# Saturation Probabilities of Continuous-Time Sigmoidal Networks

Randall D. Beer
Cognitive Science Program
Dept. of Computer Science
Dept. of Informatics
Indiana University

Bryan Daniels
Dept. of Physics
Cornell University

## Abstract

From genetic regulatory networks to nervous systems, the interactions between elements in biological networks often take a sigmoidal or S-shaped form. This paper develops a probabilistic characterization of the parameter space of continuous-time sigmoidal networks (CTSNs), a simple but dynamically-universal model of such interactions. We describe an efficient and accurate method for calculating the probability of observing effectively $M$-dimensional dynamics in an $N$-element CTSN, as well as a closed-form but approximate method. We then study the dependence of this probability on $N$, $M$, and the parameter ranges over which sampling occurs. This analysis provides insight into the overall structure of CTSN parameter space.

Please address all correspondence to:

Randall D. Beer
Cognitive Science Program
1910 E. 10$^{th}$ St. – 840 Eigenmann
Indiana University
Bloomington, IN 47406

Phone: (812) 856-0873
Fax: (812) 855-1086
Email: rdbeer@indiana.edu
URL: http://mypage.iu.edu/~rdbeer/

# 1. Introduction

A very general feature of biological networks is that the interactions between elements often take the form of a sigmoidal or S-shaped curve. Biochemical reaction kinetics [1], regulatory interactions in genetic networks [2-4], gate activation in voltage-gated ion channels [5], signaling networks [6-9], the mean firing rate of nerve cells [5], the synaptic interactions of nonspiking neurons [10], predation between species [11] (Ludwig, Jones & Holling, 1978), etc. all involve sigmoidal interactions. Boolean networks [12, 13] and continuous-time switching networks [14, 15] are both widely-used models that attempt to abstract this general character of biological networks. However, although these models have led to many important insights, Boolean networks abstract away the details of temporal patterning and both models idealize graded interactions between elements as threshold functions. Yet both temporal patterning and graded interactions play key roles in many biological networks. For this reason, we study continuous-time sigmoidal networks (CTSNs).

CTSNs are defined by the vector differential equation

$$\boldsymbol{\tau}\dot{\mathbf{y}} = -\mathbf{y} + \mathbf{W}\,\xi(\mathbf{y}+\boldsymbol{\theta}) + \mathbf{I} \qquad (1.1)$$

where $\boldsymbol{\tau}, \dot{\mathbf{y}}, \mathbf{y}, \boldsymbol{\theta},$ and $\mathbf{I}$ are length $N$ vectors, $\mathbf{W} = \{w_{ij}\}$ is an $N \times N$ matrix, and all vector operations are performed element-wise. The coupling function $\xi(\cdot)$ is assumed to be sigmoidal in nature (smooth, monotonically increasing and bounded). Common sigmoidal functions include the logistic function, the hyperbolic tangent function, the arctangent function, the error function and the algebraic Hill functions. In all of our work, we utilize the logistic function $\sigma(x) = 1/(1+e^{-x})$. However, the quantitative results we obtain can be directly transferred to any other $\xi(\cdot)$ that is related to $\sigma(\cdot)$ by a coordinate transformation (such as tanh), and our qualitative results and the methods we develop can be applied to any sigmoidal coupling function. Note that the distinction between $I$ and $\theta$ is merely semantic; with respect to the output dynamics, only the net input $I+\theta$ matters, since Equation (1.1) can be rewritten in the form $\tau \dot{\mathbf{x}} = -\mathbf{x} + \sigma(\mathbf{W}\mathbf{x} + \mathbf{I} + \boldsymbol{\theta})$ using the substitution $\mathbf{y} \mapsto \mathbf{W}\mathbf{x} + \mathbf{I}$. Thus, without loss of generality, we set $\mathbf{I} = \mathbf{0}$. An $N$-element CTSN therefore has $N$ time constants, $N$ biases, and $N^2$ weights, giving it an $(N^2+2N)$-dimensional parameter space. Interestingly, CTSNs have been proven to be universal approximators of smooth dynamics [16-18], and thus can be interpreted mathematically as a convenient basis dynamics for modeling any dynamical system.

The standard neurobiological interpretation of this model is that $y_i$ represents the mean membrane potential of the $i^{\text{th}}$ neuron, $\sigma(\cdot)$ represents its mean firing rate, $\tau_i$ represents its membrane time constant, $\theta_i$ represents its firing threshold/bias, $I_i$ represents an external input, the weights $w_{ij, j \neq i}$ represent synaptic connections from neuron $j$ to neuron $i$, and the self-interaction $w_{ii}$ represents either a literal self-collatoral or a simple active conductance. This model can also be interpreted as representing nonspiking neurons, in which case $\sigma(\cdot)$ represents saturating nonlinearities in synaptic input [10], or as lumped models of populations of neurons [19, 20].

Equation (1.1) has also been interpreted biologically as a model of genetic regulatory or signaling networks. As a genetic regulatory network model, $y_i$ represents the transcription rate of a gene, $\sigma(\cdot)$ represents the effect of a transcription factor on the transcription rate of its target gene, $\theta_i$ represents the basal expression level, connections represent regulatory interactions between genes (with the signs of the weights distinguishing between activation and repression), and self-interactions represent autoregulation [4]. The interpretation of Equation (1.1) as a model of signaling networks is similar [6]. Indeed, the similarity of signaling networks to neural networks has been previously noted [7, 21, 22].

Given the ubiquity of such continuous-time sigmoidal networks in biology and the status of CTSNs as universal dynamics approximators, it is important to understand as much as possible about the general nature of their dynamics. In particular, we are interested in the structure of the space of all possible CTSNs. Call this space $\mathcal{C}$. Although this space is obviously infinite-dimensional, it is stratified by network size $N$. Thus, to understand the space of all possible

CTSNs, we must characterize the structure of the ($N^2 + 2N$)-dimensional parameter space $\mathcal{C}(N)$ by expressions in which $N$ appears as a free parameter.

In previous work, we studied the global structure of the local bifurcation manifolds in $\mathcal{C}(N)$ [23]. Exact expressions were derived for saddle-node and Hopf bifurcation manifolds for arbitrary $N$. Visualization of these manifolds in net input space for small networks led to the discovery of a set of extremal saddle-node bifurcation manifolds that divide parameter space into regions with dynamics of different effective dimensionality due to saturation of the sigmoidal coupling function. By "effective dimensionality", we mean the actual degrees of dynamical freedom in the network outputs once the elements that are asymptotically saturated off or on regardless of the initial state are removed from consideration. Asymptotically exact approximations to these regions (denoted $\mathcal{R}_M^N(\mathbf{W})$ for a region with $M$-dimensional dynamics in an $N$-element CTSN) were defined and their geometry and combinatorics were fully characterized for arbitrary $N$ and $M$. The definition of these regions was also extended to cases where extremal saddle-node bifurcations did not exist by relaxing the definition of saturation boundaries. These extended regions were denoted $\tilde{\mathcal{R}}_M^N(\mathbf{W})$.

Due to the geometrical complexity of the $\mathcal{R}_M^N(\mathbf{W})$ and $\tilde{\mathcal{R}}_M^N(\mathbf{W})$ regions, a probabilistic characterization would be quite useful. That is, we would like to calculate the probability $P(\mathcal{R}_M^N)$ and $P(\tilde{\mathcal{R}}_M^N)$ of encountering a region of $M$-dimensional dynamics in an $N$-element CTSN given a uniform sample over specified ranges of self-interaction, coupling weight and bias parameters. Such probabilities provide a useful summary of the overall scaling properties of $\mathcal{C}(N)$, supply estimates of the dynamical complexity of randomly-chosen CTSN under various conditions, and offer guidance for selecting the best parameter ranges over which to search CTSN parameter spaces with stochastic optimization techniques. In our previous work, we also derived methods for calculating $P(\mathcal{R}_M^N)$ and $P(\tilde{\mathcal{R}}_M^N)$. Unfortunately, these methods were prohibitively expensive and really only practical for very small $N - M$. For example, computing $P(\mathcal{R}_0^4)$ using these methods required the evaluation of over 250 million 4-dimensional integrals.

In this paper, we develop an efficient method for accurately calculating these saturation probabilities. Specifically, we derive expressions for $P(\mathcal{R}_M^N)$ and $P(\tilde{\mathcal{R}}_M^N)$ that require the evaluation of only 1-dimensional integrals. We also derive closed-form but approximate expressions. Finally, we examine how these probabilities vary with $N$ and $M$, the dependence of the form of these curves on the bias, coupling weight and self-weight sampling ranges, and the implications of this probabilistic analysis for the structure of $\mathcal{C}(N)$.

## 2. Preliminaries

A single CTSN element has a steady-state input-output curve that depends on the magnitudes of its self-weight $w_{ii}$ and bias $\theta_i$. Due to the symmetry of an $N$-element CTSN, we will henceforth drop the $i$ subscripts and write our expressions for an arbitrary element. When $w > w^* = 1/\max \xi'(x)$, the element is bistable over some range of inputs (Figure 1A); otherwise it is monostable (Figure 1B). When $w > w^*$, the left and right edges of the bistable region, $I_L(w)$ and $I_R(w)$, are given by $\xi'^{-1}(1/w) - w\,\xi(\xi'^{-1}(1/w))$, where $\xi'^{-1}(\cdot)$ is 2-valued for sigmoidal $\xi(\cdot)$ [24]. For the logistic coupling function $\sigma(\cdot)$ that we employ, $w^* = 4$ and

$$I_L(w), I_R(w) \equiv \pm 2\ln\frac{\sqrt{w}+\sqrt{w-4}}{2} - \frac{w \pm \sqrt{w(w-4)}}{2}$$

[Insert Figure 1]

The asymptotic status of any particular element of a CTSN depends on the input $J$ it receives from the other elements relative to its own bias $\theta$. This input in turn depends on the statuses of these other elements and the strengths of the coupling connections from them. An element will be saturated ON when $J + \theta > I_R(w)$ and saturated OFF when $J + \theta < I_L(w)$. We will call an element ACTIVE when $I_L(w) < J + \theta < I_R(w)$, since its asymptotic status is not dictated by its input alone, but also depends upon its internal state.

In this paper, we will be concerned not with the status of elements for any particular input, but rather their status for all possible inputs. In particular, we wish to distinguish elements that will always be asymptotically OFF, ON or ACTIVE given knowledge only of the network parameters. If we have $U$ other elements in the network saturated ON, $D$ other elements saturated OFF, and $A$ other elements ACTIVE, then the range of possible inputs to an element satisfies

$$J \in [J_{\min}, J_{\max}] = \left[\min \mathcal{I}^A + \sum_{j=1}^{U} w_j, \ \max \mathcal{I}^A + \sum_{j=1}^{U} w_j\right]$$

where $\min \mathcal{I}^A$ and $\max \mathcal{I}^A$ denote the minimum and maximum values, respectively, that the sum of the weights from all of the $A$ ACTIVE elements can take and $w_j$ denotes the weight of each incoming connection from the $U$ elements that are saturated ON. Note that we have taken advantage of the fact that $\lim_{x \to +\infty} \sigma(x) = 1$ and $\lim_{x \to -\infty} \sigma(x) = 0$. For coupling functions with a nonzero lower bound, another sum would need to be added for the $D$ elements saturated OFF.

Given the information above, the asymptotic status of an element is given by the magnitude of its bias $\theta$ relative to the lower and upper boundaries

$$\begin{aligned} l^{U,A}(w, w_1, \ldots, w_{U+A}) &\equiv I_L(w) - \max \mathcal{I}^A - \sum_{j=1}^{U} w_j \\ u^{U,A}(w, w_1, \ldots, w_{U+A}) &\equiv I_R(w) - \min \mathcal{I}^A - \sum_{j=1}^{U} w_j \end{aligned} \quad (2.1)$$

In particular, an element will be asymptotically saturated OFF if $\theta < l^{U,A}$, asymptotically saturated ON if $\theta > u^{U,A}$, and ACTIVE if $l^{U,A} \leq \theta \leq u^{U,A}$.

Saturation can also occur when an element's self-weight $w$ is below the critical value for bistability. Unfortunately, the boundaries between the OFF, ACTIVE and ON regions are no longer sharply delineated by extremal saddle-node bifurcations in this case (Figure 1B). Instead, we need to extend the definitions of $I_L(w)$ and $I_R(w)$ by defining some "reasonable" alternative saturation boundaries when $w < w^*$ [23]. The exact choice is somewhat arbitrary, but one possibility is to replace $\sigma(x)$ with its piecewise-linear approximation

$$\tilde{\sigma}(y+\theta) = \begin{cases} 0 & y < -\theta - 2 \\ \dfrac{y+\theta}{4} + \dfrac{1}{2} & -\theta - 2 \le y \le -\theta + 2 \\ 1 & y > -\theta + 2 \end{cases}$$

and then define the saturation boundaries by the points at which the linear pieces intersect (gray dots in Figure 1B), giving

$$\tilde{I}_L(w) = \begin{cases} -2 & w < 4 \\ I_L(w) & w \ge 4 \end{cases}$$

$$\tilde{I}_R(w) = \begin{cases} 2 - w & w < 4 \\ I_R(w) & w \ge 4 \end{cases}$$

which are valid for all $w$. In this paper, we focus on efficiently calculating $P(\tilde{\mathcal{R}}_M^N)$ using these extended definitions of $I_L(w)$ and $I_R(w)$. Once we can calculate $P(\tilde{\mathcal{R}}_M^N)$, $P(\mathcal{R}_M^N)$ is given simply by the product of the probability that all $w_{ii} > 4$ and the probability that all $N$ elements of such a network are ACTIVE. Assuming that $w_{\min}^{\text{self}} \le 4$ and $w_{\max}^{\text{self}} > 4$, we have

$$P\left(\mathcal{R}_M^N; \theta_{\min}, \theta_{\max}, w_{\min}, w_{\max}, w_{\min}^{\text{self}}, w_{\max}^{\text{self}}\right) = \left(\frac{w_{\max}^{\text{self}} - 4}{w_{\max}^{\text{self}} - w_{\min}^{\text{self}}}\right)^N P\left(\tilde{\mathcal{R}}_M^N; \theta_{\min}, \theta_{\max}, w_{\min}, w_{\max}, 4, w_{\max}^{\text{self}}\right) \quad (2.2)$$

From this expression, it is clear that $P(\mathcal{R}_M^N)$ will decay to 0 with $N$ due to the first factor unless $w_{\min}^{\text{self}} = 4$, in which case it will have the same form as $P(\tilde{\mathcal{R}}_M^N)$.

## 3. Overall Approach

Our goal is to calculate the probability $P(\tilde{\mathcal{R}}_M^N)$ that an $N$-element CTSN has $M$ ACTIVE elements when the biases, coupling weights, and self-weights are drawn uniformly from the ranges $[\theta_{\min}, \theta_{\max}]$, $[w_{\min}, w_{\max}]$, and $[w_{\min}^{\text{self}}, w_{\max}^{\text{self}}]$, respectively. For simplicity, we assume that $w_{\min} \le 0$ and $w_{\max} \ge 0$. From the considerations in the previous section, it should be clear that the probability that a given element is ACTIVE is just the probability that its bias $\theta$ falls between the "average" left and right boundaries of its ACTIVE region clipped to the range of allowable bias values, since we are sampling uniformly in both the biases and weights. These averages are given by

$$\left\langle \left[ u^{U,A} \right]_{\theta_{min}}^{\theta_{max}} \right\rangle = \frac{1}{V_T} \int_{w_{min}^{self}}^{w_{max}^{self}} \int_{w_{min}}^{w_{max}} \cdots \int_{w_{min}}^{w_{max}} \left[ u^{U,A}(w, w_1, \ldots, w_{U+A}) \right]_{\theta_{min}}^{\theta_{max}} dw\, dw_1 \cdots dw_{U+A}$$
$$\left\langle \left[ l^{U,A} \right]_{\theta_{min}}^{\theta_{max}} \right\rangle = \frac{1}{V_T} \int_{w_{min}^{self}}^{w_{max}^{self}} \int_{w_{min}}^{w_{max}} \cdots \int_{w_{min}}^{w_{max}} \left[ l^{U,A}(w, w_1, \ldots, w_{U+A}) \right]_{\theta_{min}}^{\theta_{max}} dw\, dw_1 \cdots dw_{U+A}$$
(3.1)

where $V_T = \left(w_{max}^{self} - w_{min}^{self}\right)\left(w_{max} - w_{min}\right)^{U+A}$ is the total integration volume, and the notation $[x]_{min}^{max}$ means to clip $x$ to the range $[min, max]$.

Since the probabilities that we seek will always be expressed as differences between $\left\langle \left[ u^{U,A} \right]_{\theta_{min}}^{\theta_{max}} \right\rangle$ and $\left\langle \left[ l^{U,A} \right]_{\theta_{min}}^{\theta_{max}} \right\rangle$ and as fractions of the allowable range of biases, it will be convenient to shift and scale these average boundaries as

$$_R^N\chi_D^U \equiv \frac{\left\langle \left[ u^{U,A} \right]_{\theta_{min}}^{\theta_{max}} \right\rangle - \theta_{min}}{\theta_{max} - \theta_{min}}$$

$$_L^N\chi_D^U \equiv \frac{\left\langle \left[ l^{U,A} \right]_{\theta_{min}}^{\theta_{max}} \right\rangle - \theta_{min}}{\theta_{max} - \theta_{min}}$$

so that they lie in the range [0, 1], where $A = N - (U + D) - 1$. Note that, although we write $_R^N\chi_D^U$ and $_L^N\chi_D^U$ for notational convenience in Section 7, their values actually depend only on $U$ and $A$. Thus, $_R^{N_i}\chi_{D_i}^U = {_R^{N_j}}\chi_{D_j}^U$ and $_L^{N_i}\chi_{D_i}^U = {_L^{N_j}}\chi_{D_j}^U$ whenever $N_i - D_i = N_j - D_j$. In order to highlight this symmetry, we will often write $_R^N\chi_D^U$ and $_L^N\chi_D^U$ as $_R\chi^{U,A}$ and $_L\chi^{U,A}$, respectively.

In previous work, we calculated $\left\langle \left[ u^{U,A} \right]_{\theta_{min}}^{\theta_{max}} \right\rangle$ and $\left\langle \left[ l^{U,A} \right]_{\theta_{min}}^{\theta_{max}} \right\rangle$ by evaluating the integrals given in Equation (3.1) explicitly [23]. This could only be done algorithmically in general, and was prohibitively computationally expensive. In this paper, we take a different approach. Since the first term in Equation (2.1) depends only on the self-weight and the other two terms depend only on the coupling weights, it will be convenient to rewrite these integrals as

$$_R\chi^{U,A} = \int_{-\infty}^{\infty} F_R(x)\rho_R^{U,A}(x)dx$$
$$_L\chi^{U,A} = \int_{-\infty}^{\infty} F_L(x)\rho_L^{U,A}(x)dx \quad (3.2)$$

where we define

$$F_R\left(x;\theta_{min},\theta_{max},w_{min}^{self},w_{max}^{self}\right) \equiv \frac{1}{A_T}\int_{w_{min}^{self}}^{w_{max}^{self}}\left[\tilde{I}_R(w)-x\right]_{\theta_{min}}^{\theta_{max}}dw$$
$$F_L\left(x;\theta_{min},\theta_{max},w_{min}^{self},w_{max}^{self}\right) \equiv \frac{1}{A_T}\int_{w_{min}^{self}}^{w_{max}^{self}}\left[\tilde{I}_L(w)-x\right]_{\theta_{min}}^{\theta_{max}}dw \quad (3.3)$$

with $A_T = (\theta_{max}-\theta_{min})(w_{max}^{self}-w_{min}^{self})$ giving the total area of integration and we define $\rho_R^{U,A}(x;w_{min},w_{max})$ and $\rho_L^{U,A}(x;w_{min},w_{max})$ to be probability distributions such that $\Delta x\, \rho_L^{U,A}(x)$ gives the probability of finding $\max \mathcal{I}^A + \sum_{j=1}^{U} w_j$ between $x$ and $x+\Delta x$ and $\Delta x\, \rho_R^{U,A}(x)$ gives the probability of finding $\min \mathcal{I}^A + \sum_{j=1}^{U} w_j$ between $x$ and $x+\Delta x$. Thus, $F_L(x)$ and $F_R(x)$ respectively give the average scaled left and right boundaries of the ACTIVE range for an arbitrary element given that its input is $x$, whereas $\rho_L^{U,A}(x)$ and $\rho_R^{U,A}(x)$ give the probability distributions of this input. Assuming that we can derive expressions for $F_R(x)$, $F_L(x)$, $\rho_R^{U,A}(x)$, and $\rho_L^{U,A}(x)$, then we can compute each required $\chi$ with just a single 1-dimensional integration.

## 4. Calculating $F_R(x)$ and $F_L(x)$

Deriving expressions for $F_R(x)$ and $F_L(x)$ involves evaluating the integrals in Equation (3.3). Because of the clipping to $[\theta_{min},\theta_{max}]$, these integrands are defined piecewise in both $w$ and $x$. Appendix A.1 shows how to split these integrals and evaluate them to obtain

$$F_R(x) = \frac{1}{A_T}\left[(\theta_{max}-\theta_{min})(A_R(x)-w_{min}^{self})+\Theta_R(B_R(x))-\Theta_R(A_R(x))-(B_R(x)-A_R(x))(x+\theta_{min})\right] \quad (4.1)$$

$$F_L(x) = \frac{1}{A_T}\left[(\theta_{max}-\theta_{min})(A_L(x)-w_{min}^{self})+\Theta_L(B_L(x))-\Theta_L(A_L(x))-(B_L(x)-A_L(x))(x+\theta_{min})\right] \quad (4.2)$$

where the functions $A_{\rm R}(x)$, $B_{\rm R}(x)$, $A_{\rm L}(x)$, $B_{\rm L}(x)$, $\Theta_{\rm R}(\cdot)$ and $\Theta_{\rm L}(\cdot)$ are defined in Appendix A.1. Sample plots of the forms of $F_{\rm R}(x)$ and $F_{\rm L}(x)$ are shown in Figure 2.

[Insert Figure 2]

## 5. Calculating $\rho_{\rm R}^{U,A}(x)$ and $\rho_{\rm L}^{U,A}(x)$

In order to calculate $\rho_{\rm R}^{U,A}(x)$ and $\rho_{\rm L}^{U,A}(x)$, we will separately calculate the distribution $\rho_1^U(x)$ of $\sum_{j=1}^{U} w_j$ when $U$ weights are drawn uniformly from $[w_{\min}, w_{\max}]$ and the distributions ${}_{\rm R}\rho_2^A(x)$ and ${}_{\rm L}\rho_2^A(x)$ of $\min \mathcal{I}^A$ and $\max \mathcal{I}^A$, respectively, when $A$ weights are drawn uniformly from $[w_{\min}, w_{\max}]$. Once $\rho_1^U(x)$, ${}_{\rm R}\rho_2^A(x)$ and ${}_{\rm L}\rho_2^A(x)$ are known, then $\rho_{\rm R}^{U,A}(x) = \rho_1^U(x) * {}_{\rm R}\rho_2^A(x)$ and $\rho_{\rm L}^{U,A}(x) = \rho_1^U(x) * {}_{\rm L}\rho_2^A(x)$, where $*$ denotes convolution, since the distribution of a sum of random variables is just the convolution of their individual distributions.

Because convolution will play such an important role in this section, it is worth briefly recalling its definition and basic properties. The convolution of two functions $f(x)$ and $g(x)$ is given by $f(x) * g(x) \equiv \int_{-\infty}^{\infty} f(v) g(x-v) dv$. Convolution is commutative, associative and distributive, and the Dirac delta function $\delta(x)$ serves as its identity element. The $n^{\rm th}$ convolution power of a function $f(x)$, denoted $[f(x)]^{*n}$, is given by $\underbrace{f(x) * f(x) * \cdots * f(x)}_{n \text{ times}}$, with $[f(x)]^{*0} \equiv \delta(x)$.

*5.1 Calculating $\rho_1^U(x)$*

The distribution $\rho_1^U(x)$ of $\sum_{j=1}^{U} w_j$ is just the $U$-wise convolution of the uniform distribution over $[w_{\min}, w_{\max}]$ with itself. The convolution of uniform distributions is a classical result, due originally to Laplace [25]. If we define the box function $B_{min}^{max}(x)$ to take on the value 1 when $min \le x \le max$ and 0 otherwise, then $\rho_1^U(x)$ can be expressed as

$$\rho_1^U(x) = \left[ \frac{1}{w_{\max} - w_{\min}} B_{w_{\min}}^{w_{\max}}(x) \right]^{*U} = \left( \frac{1}{w_{\max} - w_{\min}} \right)^U \left[ B_{w_{\min}}^{w_{\max}}(x) \right]^{*U}$$

where an expression for $\left[ B_{w_{\min}}^{w_{\max}}(x) \right]^{*U}$ is derived in Appendix A.2.

## 5.2 Calculating $_R\rho_2^A(x)$ and $_L\rho_2^A(x)$

The distribution $_R\rho_2^A(x)$ of $\min \mathcal{I}^A$ can be derived as follows. First, we recall that $\mathcal{I}^A$ always contains 0 [23], so that $\min \mathcal{I}^A \leq 0$. Now suppose that $A = 1$. As we draw weights from $[w_{\min}, w_{\max}]$, a fraction $b_R$ of them will be negative, in which case they will contribute to $\min \mathcal{I}^A$, and a fraction $(1-b_R)$ of them will be positive, in which case they will contribute 0 to $\min \mathcal{I}^A$. Thus, the distribution we seek is a weighted combination of a uniform distribution from $[w_{\min}, 0]$ and a delta distribution at 0. We can write this distribution as $\frac{b_R}{|w_{\min}|} B^0_{w_{\min}}(x) + (1-b_R)\delta(x)$, with $b_R \equiv -w_{\min}/(w_{\max} - w_{\min})$ if we assume that $w_{\min}$ is always negative.

Because $\min \mathcal{I}^A$ is a sum of elements drawn from this $A = 1$ distribution, the distribution for arbitrary $A$ is just the $A$-wise convolution $\left[\frac{b_R}{|w_{\min}|} B^0_{w_{\min}}(x) + (1-b_R)\delta(x)\right]^{*A}$. Since $\delta(x)$ serves as the identity element for convolution, this expression has the same form as the polynomial $(y + (1-b_R))^A$, so using the binomial theorem it can be written as

$$_R\rho_2^A(x) = \left[\frac{b_R}{|w_{\min}|} B^0_{w_{\min}}(x) + (1-b_R)\delta(x)\right]^{*A}$$

$$= \sum_{j=0}^{A} \binom{A}{j}(1-b_R)^j \left[\frac{b_R}{|w_{\min}|} B^0_{w_{\min}}(x)\right]^{*(A-j)}$$

$$= \sum_{j=0}^{A} \binom{A}{j}(1-b_R)^j \left(\frac{b_R}{|w_{\min}|}\right)^{A-j} \left[B^0_{w_{\min}}(x)\right]^{*(A-j)}$$

By a similar derivation

$$_L\rho_2^A(x) = \sum_{j=0}^{A} \binom{A}{j}(1-b_L)^j \left(\frac{b_L}{|w_{\max}|}\right)^{A-j} \left[B^{w_{\max}}_0(x)\right]^{*(A-j)}$$

where $b_L$ is the fraction of weights that are positive, with $b_L \equiv w_{\max}/(w_{\max} - w_{\min})$ if we assume that $w_{\max}$ is always positive.

## 5.3 Putting It All Together

The distribution $\rho_R^{U,A}(x)$ of $\min \mathcal{I}^A + \sum_{j=1}^{U} w_j$ is just the convolution of the distributions $\rho_1^U(x)$ of $\sum_{j=1}^{U} w_j$ and $_R\rho_2^A(x)$ of $\min \mathcal{I}^A$. Using the expressions for $\rho_1^U(x)$ and $_R\rho_2^A(x)$ derived above, we have

$$\rho_R^{U,A}(x) = \rho_1^U(x) * {_R\rho_2^A(x)}$$

$$= \left[\left(\frac{1}{w_{max} - w_{min}}\right)^U \left[B_{w_{min}}^{w_{max}}(x)\right]^{*U}\right] * \left[\sum_{j=0}^{A}\binom{A}{j}(1-b_R)^j \left(\frac{b_R}{|w_{min}|}\right)^{A-j} \left[B_{w_{min}}^{0}(x)\right]^{*(A-j)}\right]$$

$$= \left(\frac{1}{w_{max} - w_{min}}\right)^U \left(\left[B_{w_{min}}^{w_{max}}(x)\right]^{*U} * \sum_{j=0}^{A}\binom{A}{j}(1-b_R)^j \left(\frac{b_R}{|w_{min}|}\right)^{A-j} \left[B_{w_{min}}^{0}(x)\right]^{*(A-j)}\right)$$

$$= \left(\frac{1}{w_{max} - w_{min}}\right)^U \sum_{j=0}^{A}\binom{A}{j}(1-b_R)^j \left(\frac{b_R}{|w_{min}|}\right)^{A-j} \left[B_{w_{min}}^{w_{max}}(x)\right]^{*U} * \left[B_{w_{min}}^{0}(x)\right]^{*(A-j)}$$

If we define $_R\Lambda_m^n(x) \equiv \left[B_{w_{min}}^{w_{max}}(x)\right]^{*n} * \left[B_{w_{min}}^{0}(x)\right]^{*m}$ (see Appendix A.3 for a derivation), then

$$\rho_R^{U,A}(x) = \left(\frac{1}{w_{max} - w_{min}}\right)^U \sum_{j=0}^{A}\binom{A}{j}(1-b_R)^j \left(\frac{b_R}{|w_{min}|}\right)^{A-j} {_R\Lambda_{A-j}^U(x)} \qquad U, A > 0 \qquad (5.1)$$

Note that this expression only holds for nonzero $U$ and $A$. When either $U$ or $A$ is zero, $\rho_R^{U,A}(x)$ reduces to $_R\rho_2^A(x)$ or $\rho_1^U(x)$, respectively. When both $U$ and $A$ are zero, $\rho_R^{U,A}(x)$ reduces to $\delta(x)$. Representative plots of $\rho_R^{U,A}(x)$ for various values of $U$ and $A$ are shown in Figure 3.

[Insert Figure 3]

A similar derivation for $\rho_L^{U,A}(x)$ gives

$$\rho_L^{U,A}(x) = \left(\frac{1}{w_{max} - w_{min}}\right)^U \sum_{j=0}^{A}\binom{A}{j}(1-b_L)^j \left(\frac{b_L}{|w_{max}|}\right)^{A-j} {_L\Lambda_{A-j}^U(x)} \qquad U, A > 0 \qquad (5.2)$$

where $_L\Lambda_m^n(x)$ is given in Appendix A.3.

# 6. Calculating $_R\chi^{U,A}$ and $_L\chi^{U,A}$

Armed with the above expressions for $F_R(x)$, $F_L(x)$, $\rho_R^{U,A}(x)$ and $\rho_L^{U,A}(x)$, it remains only to evaluate the 1-dimensional integrals in Equation (3.2) in order to calculate $_R\chi^{U,A}$ and $_L\chi^{U,A}$. Unfortunately, these integrals must be evaluated numerically. We show in Appendix A.5 how the integration ranges can be restricted from $\pm\infty$ because $\rho_R^{U,A}(x)$ and $\rho_L^{U,A}(x)$ are only nonzero over a limited range. We also show how various special cases can be taken into account, giving

$$_R\chi^{U,A} = \begin{cases} F_R(0) & U = A = 0 \\ (1-b_R)^A F_R(0) + \int_{(U+A)w_{min}}^{U w_{max}} F_R(x) \left[ \sum_{j=0}^{A-1} \binom{A}{j} (1-b_R)^j \left(\frac{b_R}{|w_{min}|}\right)^{A-j} \left[B_{w_{min}}^0(x)\right]^{*(A-j)} \right] dx & U = 0, A > 0 \\ \int_{(U+A)w_{min}}^{U w_{max}} F_R(x) \rho_R^{U,A}(x) dx & \text{otherwise} \end{cases} \quad (6.1)$$

$$_L\chi^{U,A} = \begin{cases} F_L(0) & U = A = 0 \\ (1-b_L)^A F_L(0) + \int_{U w_{min}}^{(U+A)w_{max}} F_L(x) \left[ \sum_{j=0}^{A-1} \binom{A}{j} (1-b_L)^j \left(\frac{b_L}{|w_{max}|}\right)^{A-j} \left[B_0^{w_{max}}(x)\right]^{*(A-j)} \right] dx & U = 0, A > 0 \\ \int_{U w_{min}}^{(U+A)w_{max}} F_L(x) \rho_L^{U,A}(x) dx & \text{otherwise} \end{cases} \quad (6.2)$$

# 7. Calculating $P(\tilde{\mathcal{R}}_M^N)$

Since $^N_R\chi_D^U$ and $^N_L\chi_D^U$ give the average active range boundaries normalized to $[0,1]$, we can define the following abbreviations for the probability of a single element having each possible asymptotic status, given $U$ total elements saturated ON and $D$ total elements saturated OFF

$$^N P_D^U(\text{ACTIVE}) = \left(^N_R\chi_D^U - ^N_L\chi_D^U\right)$$
$$^N P_D^U(\text{ON}) = \left(1 - ^N_R\chi_D^{U-1}\right) \quad (7.1)$$
$$^N P_D^U(\text{OFF}) = ^N_L\chi_{D-1}^U$$

Therefore, the probability of all $N$ elements being active in an $N$-node CTSN is simply

$$P(\tilde{\mathcal{R}}_N^N) = \left[{}^N P_0^0(\text{ACTIVE})\right]^N = \left({}_R^N\chi_0^0 - {}_L^N\chi_0^0\right)^N$$

A plot comparing our theoretical expression for $P(\tilde{\mathcal{R}}_N^N)$ to an empirical sample is shown in Figure 4.

[Insert Figure 4 Here]

Calculating $P(\tilde{\mathcal{R}}_M^N)$ for general $M \neq N$ requires additional work. Although $U + D = N - M$, there are many different ways of choosing $U$ and $D$ such that they sum to $N - M$ and, for each of those, there are many different ways of choosing which particular elements are saturated ON or OFF. Taking all of this into account, we can write

$$P(\tilde{\mathcal{R}}_M^N) = \sum_{U=0}^{N-M} \binom{N}{U}\binom{N-U}{D} {}^N A_D^U \, {}^N S_D^U \tag{7.2}$$

where $D = N - U - M$, ${}^N A_D^U$ is the probability of choosing parameters for $M = N - U - D$ elements such that these $M$ elements are ACTIVE given that $U$ elements are saturated ON and $D$ elements are saturated OFF, and ${}^N S_D^U$ is the probability of choosing parameters for $U + D$ elements such that $U$ elements are saturated ON and $D$ elements are saturated OFF. Said another way, ${}^N A_D^U$ represents an $M$-dimensional fractional hypervolume in which the parameters of $M$ elements are set so as to be ACTIVE, whereas ${}^N S_D^U$ represents a $(U + D)$-dimensional fractional hypervolume in which the parameters of $U$ elements are set so as to be ON and the parameters of $D$ elements are set so as to be OFF. Multiplying these two factors thus gives the probability that we seek as an $(M + U + D = N)$-dimensional fractional hypervolume.

It is easy to see that

$${}^N A_D^U = \left[P_D^U(\text{ACTIVE})\right]^M = \left({}_R^N\chi_D^U - {}_L^N\chi_D^U\right)^M \tag{7.3}$$

where $M = N - U - D$.

$^N S_D^U$ is more difficult to calculate. $^N S_D^U$ is clearly related to $\left[^N P_D^U (\text{ON})\right]^U \left[^N P_D^U (\text{OFF})\right]^D$. However, this expression assumes that the probability of each element being in saturation is independent of the asymptotic statuses of the other elements. Unfortunately, this is not true in general. To see why, consider the various saturation regions for a 2-element CTSN with fixed self-weights and coupling weights shown in Figure 5A. Note that the lower left-hand region, in which both elements are saturated OFF, is nonconvex. This region would be denoted $^2 Q_{12}$ using the notation from [23], in which lowered indices on the right-hand side of the $Q$ symbol denote elements that are saturated OFF and raised indices on the right-hand side denote elements that are saturated ON. The probability $^2 S_2^0$ of this region can be written as

$$^2 S_2^0 = \left[^2 P_2^0 (\text{OFF})\right]^2 - \left[^2 P_2^0 (\text{OFF}) \cap\, ^2 P_0^0 (\text{ACTIVE})\right]^2 \,^2 S_0^0$$

where $\left[^2 P_2^0 (\text{OFF})\right]^2$ gives the fractional hypervolume of the bounding box of $^2 Q_{12}$ and $\left[^2 P_2^0 (\text{OFF}) \cap\, ^2 P_0^0 (\text{ACTIVE})\right]^2 \,^2 S_0^0$ gives the fractional hypervolume of the overlap region between $^2 Q_{12}$ and $^2 Q$.

[Insert Figure 5 Here]

Likewise, for 3-element CTSNs, the region $^3 Q_{123}$ can be nonconvex at one corner and along three adjacent edges due to intersections of its bounding box with $^3 Q$ and $^3 Q_1$, $^3 Q_2$, $^3 Q_3$, respectively (Figure 5B). Thus, we must subtract off both the probability of overlap between $^3 P_3^0 (\text{OFF})$ and $^3 P_0^0 (\text{ACTIVE})$ and three times the probability of overlap between $^3 P_3^0 (\text{OFF})$ and $^3 P_1^0 (\text{ACTIVE})$ from the bounding box probability $\left[^3 P_3^0 (\text{OFF})\right]^3$, giving

$$^3 S_3^0 = \left[^3 P_3^0 (\text{OFF})\right]^3 - \left[^3 P_3^0 (\text{OFF}) \cap\, ^3 P_0^0 (\text{ACTIVE})\right]^3 \,^3 S_0^0 - 3 \left[^3 P_3^0 (\text{OFF}) \cap\, ^3 P_1^0 (\text{ACTIVE})\right]^2 \,^3 S_1^0$$

In general, the bounding box probability $\left[^N P_D^U (\text{ON})\right]^U \left[^N P_D^U (\text{OFF})\right]^D$ will overestimate $^N S_D^U$ due to the possibility of overlap with other regions in which some subset of the elements assumed saturated are actually ACTIVE. Define $\alpha$ to be the number of such unexpectedly

ACTIVE elements. We can have overlap contributions from regions with $\alpha$ between 2 and $U + D$ ACTIVE elements.

For each such region, we need to account for all of the different ways in which overlap can occur. Some number $i$ of the elements that we assumed to be saturated ON will actually be ACTIVE, in which case the actual number of ON elements will be $U - i$. In addition, for a given $i$, $\alpha - i$ of the elements that we assumed to be saturated OFF will actually be ACTIVE, in which case the actual number of OFF elements will be $D - (\alpha - i)$. The first case will happen with a probability given by the $i$-dimensional fractional hypervolume of the intersection $\left[ {}^N P_D^U (\text{ON}) \cap {}^N P_{D-\alpha+i}^{U-i} (\text{ACTIVE}) \right]^i$. The second case will happen with a probability given by the $(\alpha - i)$-dimensional fractional hypervolume of the intersection $\left[ {}^N P_D^U (\text{OFF}) \cap {}^N P_{D-\alpha+i}^{U-i} (\text{ACTIVE}) \right]^{\alpha - i}$. Then the total probability of the overlap region is given by the product of these two factors multiplied by the $(U + D - \alpha)$-dimensional fractional hypervolume ${}^N S_{D-\alpha+i}^{U-i}$ in which the parameters of $U - i$ elements are set so as to be ON and the parameters of $D - (\alpha - i)$ elements are set so as to be OFF, thus giving a $(U + D)$-dimensional fractional hypervolume for ${}^N S_D^U$.

Finally, we must allow $i$ to take on all possible values so that we can account for all possible overlaps. The smallest that $i$ can be is $\alpha - D$ unless $D > \alpha$, in which case $i = 0$. The largest that $i$ can be is $U$ unless $U > \alpha$, in which case $i = \alpha$. Putting this all together, we have the recurrence relation

$$
{}^N S_D^U = \left[ {}^N P_D^U (\text{ON}) \right]^U \left[ {}^N P_D^U (\text{OFF}) \right]^D - \\
\sum_{\alpha=2}^{U+D} \sum_{i=\max(\alpha-D,0)}^{\min(\alpha,U)} \binom{U}{i} \binom{D}{\alpha - i} \left[ {}^N P_D^U (\text{ON}) \cap {}^N P_{D-\alpha+i}^{U-i} (\text{ACTIVE}) \right]^i \left[ {}^N P_D^U (\text{OFF}) \cap {}^N P_{D-\alpha+i}^{U-i} (\text{ACTIVE}) \right]^{\alpha - i} {}^N S_{D-\alpha+i}^{U-i} \quad (7.4)
$$

where the basis cases are given by defining the value of any square bracketed expression with an exponent of 0 to be 1. Note that this implies that ${}^N S_0^0 = 1$.

Finally, we must rewrite the intersection probability expressions appearing in Equation (7.4) in terms of ${}^N_R \chi_D^U$ and ${}^N_L \chi_D^U$. In the case of the ON overlap factor, we can use (7.1) to write

$$^{N}P_{D}^{U}(\text{ON}) \cap {}^{N}P_{D-\alpha+i}^{U-i}(\text{ACTIVE}) = \left[{}_{R}^{N}\chi_{D}^{U-1}, 1\right] \cap \left[{}_{L}^{N}\chi_{D-\alpha+i}^{U-i}, {}_{R}^{N}\chi_{D-\alpha+i}^{U-i}\right]$$

It can be shown that the following relationships hold among the $\chi$s

$$_{L}^{N}\chi_{j}^{i} \leq {}_{R}^{N}\chi_{j}^{i}; \tag{7.5}$$

$$_{R}^{N}\chi_{j}^{i} \geq {}_{R}^{N}\chi_{J}^{i}, \quad {}_{R}^{N}\chi_{j}^{i} \geq {}_{R}^{N}\chi_{j}^{I}; \tag{7.6}$$

$$_{L}^{N}\chi_{j}^{i} \leq {}_{L}^{N}\chi_{J}^{i}, \quad {}_{L}^{N}\chi_{j}^{i} \leq {}_{L}^{N}\chi_{j}^{I}; \tag{7.7}$$

for all $i, j, I, J$ where $I > i$ and $J > j$. Equation (7.5) follows from the fact that $l^{U,A} \leq u^{U,A}$ in Equation (2.1) because $I_{L}(w) \leq I_{R}(w)$ and $\min \mathcal{I}^{A} \leq \max \mathcal{I}^{A}$. The other relations can be derived by analyzing what happens to $\min \mathcal{I}^{A}$ and $\max \mathcal{I}^{A}$ when we increase $i$ or $j$ (and hence decrease $A$) by one.

We can use these relations to write expressions for the probabilities of the overlap regions. For example, we know from Equation (7.5) that the lower bound of $\left[{}_{R}^{N}\chi_{D}^{U-1}, 1\right] \cap \left[{}_{L}^{N}\chi_{D-\alpha+i}^{U-i}, {}_{R}^{N}\chi_{D-\alpha+i}^{U-i}\right]$ must be the larger of ${}_{R}^{N}\chi_{D}^{U-1}$ and ${}_{L}^{N}\chi_{D-\alpha+i}^{U-i}$ and, by Equation (7.7), we can conclude that it must be ${}_{R}^{N}\chi_{D}^{U-1}$. Likewise, we know that the upper bound of the intersection must be the smaller of 1 and ${}_{R}^{N}\chi_{D-\alpha+i}^{U-i}$ and, since ${}_{R}^{N}\chi_{j}^{i} \leq 1$ by definition, it must be ${}_{R}^{N}\chi_{D-\alpha+i}^{U-i}$. Finally, we know that $D - (\alpha - i) \leq D$ (because $0 \leq i \leq \alpha$ and $2 \leq \alpha \leq U + D$) and $U - i \leq U - 1$ (because if $i = 0$, a basis case applies in Equation (7.4) and ${}^{N}P_{D}^{U}(\text{ON}) \cap {}^{N}P_{D-\alpha+i}^{U-i}(\text{ACTIVE})$ will never be evaluated). Thus, by Equation (7.6), ${}_{R}^{N}\chi_{D}^{U-1} \leq {}_{R}^{N}\chi_{D-\alpha+i}^{U-i}$ and the two regions do in fact overlap, allowing us to write

$$^{N}P_{D}^{U}(\text{ON}) \cap {}^{N}P_{D-\alpha+i}^{U-i}(\text{ACTIVE}) = \left({}_{R}^{N}\chi_{D-\alpha+i}^{U-i} - {}_{R}^{N}\chi_{D}^{U-1}\right)$$

By analogous arguments, the OFF overlap in Equation (7.4) can be expressed as

$$^{N}P_{D}^{U}(\text{OFF}) \cap {}^{N}P_{D-\alpha+i}^{U-i}(\text{ACTIVE}) = \left({}_{L}^{N}\chi_{D-1}^{U} - {}_{L}^{N}\chi_{D-\alpha+i}^{U-i}\right)$$

Substituting these expressions for the overlap probabilities in Equation (7.4), we finally obtain

$$^N S_D^U = \left[1 - {}_R^N\chi_D^{U-1}\right]^U \left[{}_L^N\chi_{D-1}^U\right]^D - \sum_{\alpha=2}^{U+D} \sum_{i=\max(\alpha-D,0)}^{\min(\alpha,U)} \binom{U}{i}\binom{D}{\alpha-i}\left[{}_R^N\chi_{D-\alpha+i}^{U-i} - {}_R^N\chi_D^{U-1}\right]^i \left[{}_L^N\chi_{D-1}^U - {}_L^N\chi_{D-\alpha+i}^{U-i}\right]^{\alpha-i} {}^N S_{D-\alpha+i}^{U-i} \quad (7.8)$$

In summary, Equation (7.2), combined with Equations (7.3), (7.8), (6.1), (6.2), (5.1), (5.2), (4.1), (4.2) and the results derived in the Appendix, provides an efficient method for evaluating $P(\tilde{\mathcal{R}}_M^N)$ and, by Equation (2.2), $P(\mathcal{R}_M^N)$. This method is implemented in an electronic supplement [26] distributed as a *Mathematica* notebook [27]. In order to further improve efficiency, this implementation uses memoization techniques to cache the values of ${}_R^N\chi_D^U$ and ${}_L^N\chi_D^U$ so that they only have to be computed once. With memoization, it can be shown that the total number of distinct 1-dimensional integrals that our method must evaluate in order to calculate $P(\tilde{\mathcal{R}}_M^N)$ is simply $(N-M+1)(N-M+2)$.

## 8. Closed-Form Approximations

It is possible to derive closed-form approximations to ${}_R\chi^{U,A}$ and ${}_L\chi^{U,A}$ by developing approximations to $F_R(x)$, $F_L(x)$, $\rho_R^{U,A}(x)$ and $\rho_L^{U,A}(x)$.

Approximations to $F_R(x)$ and $F_L(x)$ can be derived by replacing $I_R(w)$ and $I_L(w)$ with the piecewise linear approximations

$$\hat{I}_R(w) = \begin{cases} 2-w & w \le 4 \\ -2 & w > 4 \end{cases}$$

$$\hat{I}_L(w) = \begin{cases} -2 & w \le 4 \\ 2-w & w > 4 \end{cases}$$

By repeating the derivations given in Appendix A.1, we obtain

$$\hat{F}_R(x) = \frac{1}{A_T}\left[(\theta_{\max} - \theta_{\min})(\hat{A}_R(x) - w_{\min}^{\text{self}}) + \hat{\Theta}_R(\hat{B}_R(x)) - \hat{\Theta}_R(\hat{A}_R(x)) - (\hat{B}_R(x) - \hat{A}_R(x))(x + \theta_{\min})\right]$$

$$\hat{F}_\mathrm{L}(x) = \frac{1}{A_T}\left[(\theta_{\max} - \theta_{\min})(\hat{A}_\mathrm{L}(x) - w_{\min}^{\mathrm{self}}) + \hat{\Theta}_\mathrm{L}(\hat{B}_\mathrm{L}(x)) - \hat{\Theta}_\mathrm{L}(\hat{A}_\mathrm{L}(x)) - (\hat{B}_\mathrm{L}(x) - \hat{A}_\mathrm{L}(x))(x + \theta_{\min})\right]$$

where the functions $\hat{A}_\mathrm{R}(x)$, $\hat{B}_\mathrm{R}(x)$, $\hat{A}_\mathrm{L}(x)$, $\hat{B}_\mathrm{L}(x)$, $\hat{\Theta}_\mathrm{R}(\cdot)$ and $\hat{\Theta}_\mathrm{L}(\cdot)$ are defined in Appendix A.6. Plots of the approximations $\hat{F}_\mathrm{R}(x)$ and $\hat{F}_\mathrm{L}(x)$ (dashed gray curves) are compared to $F_\mathrm{R}(x)$ and $F_\mathrm{L}(x)$ (black curves), respectively in Figure 2.

Normal approximations to $\rho_\mathrm{R}^{U,A}(x)$ and $\rho_\mathrm{L}^{U,A}(x)$ can be derived as follows. By the central limit theorem, $\rho_1^U(x)$ approaches a normal distribution for large $U$ with mean $\mu_1 = U(w_{\min} + w_{\max})/2$ and variance $\sigma_1^2 = U(w_{\max} - w_{\min})^2/12$. Likewise, for large $A$, $_\mathrm{R}\rho_2^A(x)$ approaches a normal distribution with mean

$$_\mathrm{R}\mu_2 = A\int_{-\infty}^{\infty} x\left[\frac{b_\mathrm{R}}{|w_{\min}|}B_{w_{\min}}^0(x) + (1-b_\mathrm{R})\delta(x)\right]dx$$
$$= A\int_{w_{\min}}^0 x\frac{b_\mathrm{R}}{|w_{\min}|}dx + \int_{-\infty}^{\infty} x(1-b_\mathrm{R})\delta(x)dx$$
$$= -\frac{A\, b_\mathrm{R}\, w_{\min}^2}{2|w_{\min}|}$$

and variance

$$_\mathrm{R}\sigma_2^2 = A\int_{-\infty}^{\infty}(x - {}_\mathrm{R}\mu_2)^2\left[\frac{b_\mathrm{R}}{|w_{\min}|}B_{w_{\min}}^0(x) + (1-b_\mathrm{R})\delta(x)\right]dx$$
$$= A\int_{w_{\min}}^0 (x - {}_\mathrm{R}\mu_2)^2 \frac{b_\mathrm{R}}{|w_{\min}|}dx + A\int_{-\infty}^{\infty}(x - {}_\mathrm{R}\mu_2)^2(1-b_\mathrm{R})\delta(x)dx$$
$$= \frac{A\, b_\mathrm{R}(4 - 3b_\mathrm{R})w_{\min}^2}{12}$$

Since $\rho_1^U(x)$ and $_\mathrm{R}\rho_2^A(x)$ can be approximated by normal distributions with means $\mu_1$ and $_\mathrm{R}\mu_2$ and variances $\sigma_1^2$ and $_\mathrm{R}\sigma_2^2$, respectively, for sufficiently large $U$ and $A$, $\rho_\mathrm{R}^{U,A}(x)$ can be approximated by a normal distribution with mean $\mu_1 + {}_\mathrm{R}\mu_2$ and variance $\sigma_1^2 + {}_\mathrm{R}\sigma_2^2$. Some representative plots of $\hat{\rho}_\mathrm{R}^{U,A}(x)$ for various values of $U$ and $A$ are shown as dashed gray curves in Figure 3. The same considerations apply to $\rho_\mathrm{L}^{U,A}(x)$ with

$$_L\mu_2 = \frac{A\, b_L\, w_{\max}^2}{2|w_{\max}|}$$

$$_L\sigma_2^2 = \frac{A\, b_L(4 - 3b_L)w_{\max}^2}{12}$$

Finally, in order to compute the closed-form approximations $_R\hat{\chi}^{U,A}$ and $_L\hat{\chi}^{U,A}$, we must evaluate the integrals in Equation (3.2) using $\hat{F}_R(x)$, $\hat{F}_L(x)$, $\hat{\rho}_R^{U,A}(x)$ and $\hat{\rho}_L^{U,A}(x)$. By rewriting $\hat{F}_R(x)$ and $\hat{F}_L(x)$ as piecewise polynomials in $x$ and then performing the integration with the Gaussian probability density function piecewise (see Appendix A.6), we obtain

$$_R\hat{\chi}^{U,A} = \begin{cases} \hat{F}_R(0) & U = A = 0 \\ \dfrac{1}{A_T}\left[(\theta_{\max} - \theta_{\min})\left(G_{\hat{A}_R} - w_{\min}^{\text{self}}\right) + G_{\hat{\Theta}_R(\hat{B}_R)} - G_{\hat{\Theta}_R(\hat{A}_R)} - G_{x\hat{B}_R} + G_{x\hat{A}_R} - \theta_{\min}\left(G_{\hat{B}_R} - G_{\hat{A}_R}\right)\right] & \text{otherwise} \end{cases}$$

$$_L\hat{\chi}^{U,A} = \begin{cases} \hat{F}_L(0) & U = A = 0 \\ \dfrac{1}{A_T}\left[(\theta_{\max} - \theta_{\min})\left(G_{\hat{A}_L} - w_{\min}^{\text{self}}\right) + G_{\hat{\Theta}_L(\hat{B}_L)} - G_{\hat{\Theta}_L(\hat{A}_L)} - G_{x\hat{B}_L} + G_{x\hat{A}_L} - \theta_{\min}\left(G_{\hat{B}_L} - G_{\hat{A}_L}\right)\right] & \text{otherwise} \end{cases}$$

where expressions for the various $G$ terms are given in Appendix A.6. A comparison of the resulting approximation $\hat{P}(\tilde{\mathcal{R}}_N^N)$ (dashed gray curve) with $P(\tilde{\mathcal{R}}_N^N)$ (solid curve) is shown in Figure 4.

## 9. The Form of $P(\tilde{\mathcal{R}}_M^N)$

Now that we can accurately and efficiently calculate $P(\tilde{\mathcal{R}}_M^N)$, we can begin to study its properties. Although this is not the place for a systematic study of these curves, in order to illustrate the application of the expressions derived in this paper we will briefly examine the typical shape of these curves, the dependence of this shape on the parameter ranges over which sampling occurs, and an example of the implications of this probabilistic characterization for the structure of $C(N)$.

A typical family of $P(\tilde{\mathcal{R}}_M^N)$ curves is shown in Figure 6A for nominal parameter ranges of [-10, 10]. Here we plot $P(\tilde{\mathcal{R}}_{N-k}^N)$ in black for $k$ varying from 0 to $N$ and we plot $P(\tilde{\mathcal{R}}_M^N)$ in gray for $M$ varying from 0 to $N$. $P(\tilde{\mathcal{R}}_M^N)$ and $P(\tilde{\mathcal{R}}_{N-k}^N)$ are two complementary ways of looking at the same information. $P(\tilde{\mathcal{R}}_M^N)$ are the curves of $M$-dimensional dynamics, while $P(\tilde{\mathcal{R}}_{N-k}^N)$ are the curves of $k$-codimensional dynamics. $P(\tilde{\mathcal{R}}_M^N)$ is most appropriate when the focus is on the dynamics of a given dimensionality (e.g., $P(\tilde{\mathcal{R}}_2^N)$ gives the probability of 2-dimensional dynamics as a function of $N$). In contrast, $P(\tilde{\mathcal{R}}_{N-k}^N)$ is most appropriate when the focus is on the corresponding combinatorial structure across $N$ (e.g., $P(\tilde{\mathcal{R}}_{N-1}^N)$ always gives the probability of the "poles" sprouting from the central hypercube for all $N$). Note that every $P(\tilde{\mathcal{R}}_{N-k}^N)$ curve begins on the curve $P(\tilde{\mathcal{R}}_0^N)$ because $P(\tilde{\mathcal{R}}_{N-k}^N)$ only exists for $k \leq N$ and thus first appears at $N = k$. Likewise, every $P(\tilde{\mathcal{R}}_M^N)$ curve begins on the curve $P(\tilde{\mathcal{R}}_N^N)$ because $P(\tilde{\mathcal{R}}_M^N)$ only exists for $M \leq N$ and thus first appears at $N = M$.

The most striking feature of this plot is that $P(\tilde{\mathcal{R}}_N^N)$ eventually grows to 1 while all other curves eventually fall to 0. Thus, as $N$ grows, the probability of all elements being ACTIVE (in the extended sense of $\tilde{\mathcal{R}}_M^N$) approaches 1. However, $P(\tilde{\mathcal{R}}_N^N)$ exhibits an initial dip before growing to 1. Another interesting feature is that, although $P(\tilde{\mathcal{R}}_0^N)$ (the probability that all elements are saturated) decreases monotonically to 0, some intermediate curves temporarily flatten before falling to 0. Indeed, $P(\tilde{\mathcal{R}}_{N-1}^N)$ shows evidence of at least one inflection point.

[Insert Figure 6 Here]

How do these curves vary with the parameter ranges over which sampling occurs? We will examine only variations of $[\theta_{min}, \theta_{max}]$ and $[w_{min}, w_{max}]$ that are symmetric about 0 here. In Figure 6B and 6C we plot $P(\tilde{\mathcal{R}}_{N-k}^N)$ and $P(\tilde{\mathcal{R}}_M^N)$ for $\theta$ ranges of half and double the nominal size, respectively. Note that we have the same asymptotic behavior as before: $P(\tilde{\mathcal{R}}_N^N)$ grows to 1 and all other curves fall to 0 in both cases. However, for the smaller bias range (Figure 6B), $P(\tilde{\mathcal{R}}_N^N)$ dips less and rises much more quickly to 1 and all other curves fall much more quickly to 0. In contrast, for the larger $\theta$ range (Figure 6C), $P(\tilde{\mathcal{R}}_N^N)$ dips much more and rises more slowly to 1. Some of the other curves also exhibit clear peaks, especially $P(\tilde{\mathcal{R}}_{N-1}^N)$ and $P(\tilde{\mathcal{R}}_1^N)$, $P( \tilde{\mathcal{R}}_2^N )$, $P( \tilde{\mathcal{R}}_3^N )$, $P( \tilde{\mathcal{R}}_4^N )$ and $P( \tilde{\mathcal{R}}_5^N )$. Figure 6C, for example, shows that the probability of observing 2-dimensional dynamics is higher in 4-element networks than it is in 2-element networks for these parameter ranges.

Varying coupling weight ranges produces the opposite effect, with the smaller weight range (Figure 6D) producing effects similar to the larger bias range (Figure 6C) and the larger weight range (Figure 6E) producing effects similar to the smaller bias range (Figure 6B). These observations suggest a tradeoff between multiple growth processes. Although the process leading to the growth of $P( \tilde{\mathcal{R}}_N^N )$ always eventually dominates, other processes must oppose this growth, allowing other curves representing CTSNs with some number of saturated elements to temporarily dominate. Apparently, the relative tradeoff between these growth processes depends on the bias range and the coupling weight range in opposite ways. This is exactly the sort of insight into the structure of *C*(*N*) that we would like to be able to identify and explain using the theory developed in this paper.

Let us begin with the question of why $P( \tilde{\mathcal{R}}_N^N )$ always eventually approaches 1. For this to happen, it must be increasingly likely that an element's bias range falls entirely between the lower and upper bounds set by $l^{U,A}$ and $u^{U,A}$ in Equation (2.1) as *N* increases. Intuitively, this occurs because the range of possible inputs received by a given element increases with *N*. We can characterize this growth quantitatively by calculating the mean locations of these upper and lower boundaries (see Appendix A.7):

$$\langle u \rangle(N) = \frac{\Theta_R \left( w_{max}^{self} \right) - \Theta_R \left( w_{min}^{self} \right)}{w_{max}^{self} - w_{min}^{self}} + \sum_{A=0}^{N-1} \frac{A w_{min}^2}{2 \left( w_{max} - w_{min} \right)} P( \tilde{\mathcal{R}}_A^{N-1} )$$

$$\langle l \rangle(N) = \frac{\Theta_L \left( w_{max}^{self} \right) - \Theta_L \left( w_{min}^{self} \right)}{w_{max}^{self} - w_{min}^{self}} - \sum_{A=0}^{N-1} \frac{A w_{max}^2}{2 \left( w_{max} - w_{min} \right)} P( \tilde{\mathcal{R}}_A^{N-1} )$$

Plots of these expressions for an arbitrary element in an *N*-element CTSN are shown in Figure 7 for the same five parameter ranges shown in Figure 6. In all cases, it is clear that $\langle l \rangle(N)$ and $\langle u \rangle(N)$ will eventually exceed any fixed bias range. For smaller bias ranges, these curves grow more quickly and therefore exceed the smaller allowable range even for relatively small $N$ (Figure 7B), resulting in an earlier dominance of $P(\tilde{\mathcal{R}}_N^N)$ (Figure 6B). In contrast, for larger bias ranges, these curves grow more slowly and therefore exceed the larger allowable range only for larger $N$ (Figure 7C), resulting in a later dominance of $P(\tilde{\mathcal{R}}_N^N)$ (Figure 6C). The opposite effect occurs for the coupling weight parameter ranges, with smaller ranges causing the mean boundaries to grow more slowly (Figure 7D) and larger ranges causing the mean boundaries to grow more quickly (Figure 7E). Note that the boundary curves appear to approach straight lines as $N$ increases. This is because $P(\tilde{\mathcal{R}}_A^{N-1})$ approaches 0 for all $A$ except $A = N-1$, which approaches 1, giving the linear expressions

$$\langle u \rangle(N) \approx \frac{\Theta_R\left(w_{\max}^{\text{self}}\right) - \Theta_R\left(w_{\min}^{\text{self}}\right)}{w_{\max}^{\text{self}} - w_{\min}^{\text{self}}} + \frac{w_{\min}^2}{2(w_{\max} - w_{\min})}(N-1)$$

$$\langle l \rangle(N) \approx \frac{\Theta_L\left(w_{\max}^{\text{self}}\right) - \Theta_L\left(w_{\min}^{\text{self}}\right)}{w_{\max}^{\text{self}} - w_{\min}^{\text{self}}} - \frac{w_{\max}^2}{2(w_{\max} - w_{\min})}(N-1)$$

for sufficiently large $N$.

[Insert Figure 7 Here]

How can we explain the forms of the $P(\tilde{\mathcal{R}}_{N-k}^N)$ curves before the asymptotic growth of $P(\tilde{\mathcal{R}}_N^N)$ dominates? Consider, for example, the $P(\tilde{\mathcal{R}}_{N-1}^N)$ curve in Figure 6C, which begins near 0.9, drops to almost 0.1, then rises above 0.2 before finally dropping toward 0. If we directly expand Equation (7.2) for $P(\tilde{\mathcal{R}}_{N-1}^N)$, we obtain

$$P(\tilde{\mathcal{R}}_{N-1}^N) = N \, {}_L^N\chi_0^0 \left({}_R^N\chi_1^0 - {}_L^N\chi_1^0\right)^{N-1} + N\left(1 - {}_R^N\chi_0^0\right)\left({}_R^N\chi_0^1 - {}_L^N\chi_0^1\right)^{N-1} \tag{7.9}$$

How should we interpret this expression? The two terms come from the fact that the single element that is saturated in $P(\tilde{\mathcal{R}}_{N-1}^N)$ can be saturated either OFF (the first term) or ON (the second term). Each term is itself composed of three factors. The first factor is combinatorial; there are *N* choices for which element is in saturation. The second factor can be interpreted as the mean normalized length of each region of saturation. Finally, the third factor can be interpreted as the mean normalized width of each region of saturation raised to the (*N*-1)th power to account for the *N*-1 active elements. The light gray rectangular regions in Figure 5A, which correspond to the four ways that a single element can be saturated in a 2-element network, illustrates these three factors.

Figure 8 illustrates how the interaction of these three factors in the two terms of Equation (7.9) combine to produce the form of the curve $P(\tilde{\mathcal{R}}_{N-1}^{N})$ seen in Figure 6C. Figure 8A shows how the mean normalized length (solid gray curve) of a region contributing to $P(\tilde{\mathcal{R}}_{N-1}^{N})$ decreases with *N* as the central region $^{N}\mathcal{Q}$ grows, while the mean normalized width (dashed black curve) grows with *N* to include the entire available bias range. Figure 8A also shows how the mean normalized width raised to the (*N*-1)th power (solid black curve), which is the actual factor that appears in Equation (7.9), dips and then rises. Figure 8B shows that the product of the last two factors in the second term of Equation (7.9) drops quickly toward zero (dotted black curve). However, in order to complete the second term in Equation (7.9) we must multiply by *N*, resulting in a curve that drops, rises and then drops again (dashed black curve). The first term in Equation (7.9) takes a similar form (dashed gray curve). The sum of these two terms then gives the form of $P(\tilde{\mathcal{R}}_{N-1}^{N})$ (solid black curve) that we see in Figure 6C. Decompositions of the forms of the other curves can be performed in a similar manner; however they become more complicated due to the overlap considerations described in Section 7 and illustrated in Figure 5.

[Insert Figure 8 Here]

## 10. Conclusions

The importance of understanding both the local and global structure of the parameter spaces of biological models is becoming increasingly recognized across many subfields of biology. In neuroscience, attempts to fit neural models to data and to understand the impact of various properties of nerve cells on their activity patterns have forced a consideration of the parameter space structure of Hodgkin-Huxley models [28-30]. In addition, the model parameter space structure of experimentally well-characterized neural circuits, such as the crustacean stomatogastric ganglion, has been studied in some detail [31], with implications for both averaging of neuronal measurements of identified cells across multiple animals [32] and neuromodulation [33]. There has also been recent interest in the parameter space structure of signaling networks, with applications to the distribution of bistable switches in biochemical reaction spaces [34] and of genetic regulatory networks, with applications to the robustness of the Drosophila segment polarity gene network [35]. Finally, there is growing evidence for universal patterns in the distribution of loosely and tightly constrained directions in the model parameter spaces that arise across systems biology [36]. Given the tremendous variability of biological systems and the rate of accumulation of new data, it seems likely that characterizing the structure of the model parameter spaces that arise in systems biology will only grow in importance in the future.

Unfortunately, we have very little general theoretical understanding of either the local or global structure of such high-dimensional parameter spaces of dynamical systems. The work described in this paper is part of a larger program [23] to characterize the parameter space structure of an important class of models that captures the sigmoidally-coupled nature of many biological networks. Previous work has explored the central role that saturation plays in organizing the overall structure of the space $C(N)$ of all possible continuous-time sigmoidal networks (CTSNs), characterizing the global structure of the local bifurcation manifolds of $C(N)$ and of an asymptotically exact approximation to the extremal saddle-node bifurcation manifolds that divide $C(N)$ into regions with dynamics of different effective dimensionality. This paper has focused on a probabilistic characterization of these regions of CTSN parameter space. Specifically, we have developed an efficient exact method for calculating saturation probabilities of CTSNs of arbitrary size involving the evaluation of only 1-dimensional integrals, described a closed-form approximation that can be evaluated directly, explored the shapes of the resulting saturation probability curves and their dependence on parameter ranges, and examined how the shapes of these saturation probability curves relate to the combinatorial and geometric structure of $\mathcal{C}(N)$.

Although we have carried out our analysis for a specific sigmoidal coupling function, the logistic function, our results can be straightforwardly translated to any other coupling function related to the logistic function by a coordinate transformation. Furthermore, much of our analysis can be applied to a much wider class of sigmoidally-coupled models. Because of the way we have split up the calculations in Equation (3.2), only the calculation of $F_R(x)$ and $F_L(x)$ in Section 4 depend on the details of the actual sigmoidal function used (specifically, the functions $I_L(w)$ and $I_R(w)$ will be different for different sigmoidal functions). The calculation of $\rho_R^{U,A}(x)$ and $\rho_L^{U,A}$ in Section 5 and of the integration ranges for ${}_R^N\chi_D^U$ and ${}_L^N\chi_D^U$ in Section 6 depend only on the additive linear nature of the interactions and the upper and lower bounds of the sigmoidal function utilized (because $u^{U,A}$ and $l^{U,A}$ in Equation (2.1) depend on these bounds). The calculations in Section 7 are independent of the particular sigmoidal function chosen.

There are a number of future directions in which this work could be taken. In terms of the general program to characterize the structure of $\mathcal{C}(N)$, we would like to understand the global structure of nonextremal codimension-1 local bifurcation manifolds, higher codimension local bifurcation manifolds, and global bifurcation manifolds. Regarding the particular probabilistic characterization described in this paper, it would be useful to undertake a more detailed study of the forms of $P(\tilde{\mathcal{R}}_M^N)$ and their dependence on parameter ranges. For example, it would be interesting to examine the scaling processes underlying the forms of other saturation probability curves as we did for $P(\tilde{\mathcal{R}}_{N-1}^N)$ in Section 9. It would also be useful to derive simpler closed-form approximations for $P(\tilde{\mathcal{R}}_M^N)$. Finally, building on the preliminary work in [23], we could derive approximations for the probabilities of specific phase portraits.

Perhaps the most interesting direction for future extension of the work described in this paper would be to systematically study the impact of network architecture on the probabilities of dynamics of given effective dimensionality. The architecture of biological networks and its relationship to network dynamics is a topic of growing general interest in biology [37-41]. Since a CTSN architecture simply corresponds to a subspace of $\mathcal{C}(N)$ defined by some subset of coupling weights being set to 0, we can derive an expression for $P\left(\tilde{\mathcal{R}}_M^N\right)$ for any given architecture by simply restricting our integration to the corresponding subspace. However, deriving a general architecture-dependent expression for $P\left(\tilde{\mathcal{R}}_M^N\right)$ would require generalizing the permutation symmetry of a fully-connected network, which we have relied on to simplify our calculations, to the arbitrary symmetry groups characterizing possible architectures over an $N$-element CTSN.

# Appendix

*A.1 Derivation of $F_R(x)$ and $F_L(x)$*

Since the integrands in Equation (3.3) are defined piecewise in both $w$ and $x$, we need to split the integrals accordingly. For $\tilde{I}_R(w) - x$ in "general position", we can identify two key points at which the integral must be split (Figure A1A). Point $A_R$ is defined to be the $w$ value at which $\tilde{I}_R(w) - x$ intersects $\theta_{max}$. Since this value may lie outside the interval $[w_{min}^{self}, w_{max}^{self}]$, $A_R$ must be clipped to that range. Likewise, point $B_R$ is defined to be the $w$ value at which $\tilde{I}_R(w) - x$ intersects $\theta_{min}$. Since this value may also lie outside the interval $[w_{min}^{self}, w_{max}^{self}]$, it must also be clipped to that range. Note that, due to the clipping, one or both of these points may coincide with another point or with the limits of integration. With these two points defined, we can easily write $F_R(x)$ as

[Insert Figure A1]

$$F_R(x) = \frac{1}{A_T}\left[\int_{w_{min}^{self}}^{A_R(x)}(\theta_{max} - \theta_{min})dw + \int_{A_R(x)}^{B_R(x)}(\tilde{I}_R(w) - x - \theta_{min})dw\right]$$

where

$$A_R(x) \equiv \left[\tilde{I}_R^{-1}(x + \theta_{max})\right]_{w_{min}^{self}}^{w_{max}^{self}} \qquad A_L(x) \equiv \left[\tilde{I}_L^{-1}(x + \theta_{max})\right]_{w_{min}^{self}}^{w_{max}^{self}}$$

$$B_R(x) \equiv \left[\tilde{I}_R^{-1}(x + \theta_{min})\right]_{w_{min}^{self}}^{w_{max}^{self}} \qquad B_L(x) \equiv \left[\tilde{I}_L^{-1}(x + \theta_{min})\right]_{w_{min}^{self}}^{w_{max}^{self}}$$

The above expressions are written in terms of $\tilde{I}_R^{-1}(y)$ and $\tilde{I}_L^{-1}(y)$, respectively. Unfortunately, although the linear parts of $\tilde{I}_R(w)$ and $\tilde{I}_L(w)$ are easily invertible (if we define $\tilde{I}_L^{-1}(y) \equiv -\infty$ when $y \geq -2$), the nonlinear parts cannot be algebraically inverted. Therefore, the inverse of these nonlinear parts are numerically approximated by cubic interpolation on a fine mesh of points.

It is straightforward to evaluate these integrals indefinitely and then substitute the given limits of integration. If we define

$$\Theta_R(w) \equiv \int \tilde{I}_R(w)\,dw$$

$$= \begin{cases} 2w - \dfrac{w^2}{2} & w \le 4 \\ -(2w-2)\ln\left(\sqrt{w}+\sqrt{w-4}\right)+\left(\dfrac{w}{2}+\dfrac{1}{2}\right)\sqrt{w(w-4)}-\dfrac{w^2}{2}+(w-1)\ln 4 + 4 & w > 4 \end{cases}$$

then we obtain Equation (4.1).

A similar derivation can be applied for $F_L(x)$, with $A_L(x)$ and $B_L(x)$ defined similarly (Figure A1B), giving

$$F_L(x) = \dfrac{1}{A_T}\left[\int_{w_{min}^{self}}^{A_L(x)}(\theta_{max}-\theta_{min})\,dw + \int_{A_L(x)}^{B_L(x)}(\tilde{I}_L(w)-x-\theta_{min})\,dw\right]$$

If we evaluate these integrals indefinitely, substitute the given limits of integration, and define

$$\Theta_L(w) \equiv \int \tilde{I}_L(w)\,dw$$

$$= \begin{cases} -2w & w \le 4 \\ (2w-2)\ln\left(\sqrt{w}+\sqrt{w-4}\right)-\left(\dfrac{w}{2}+\dfrac{1}{2}\right)\sqrt{w(w-4)}-\dfrac{w^2}{2}-(w-1)\ln 4 - 4 & w > 4 \end{cases}$$

then we obtain Equation (4.2).

*A.2 Derivation of $\left[B_{min}^{max}(x)\right]^{*n}$*

Using the inverse Fourier transform of the $n^{th}$ power of the characteristic function of the uniform distribution over [0,1] [25, 42], the convolution power of the box function $B_0^1(x)$ can be expressed as

$$\left[B_0^1(x)\right]^{*n} = \dfrac{1}{2(n-1)!}\sum_{k=0}^{n}(-1)^k\binom{n}{k}(x-k)^{n-1}\,\mathrm{sgn}(x-k)$$

This expression can easily be generalized to arbitrary *min* and *max* as

$$\left[B_{min}^{max}(x)\right]^{*n} = \frac{1}{2(n-1)!}\sum_{k=0}^{n}(-1)^k\binom{n}{k}(x - k\,max - (n-k)\,min)^{n-1}\,\text{sgn}(x - k\,max - (n-k)\,min)$$

with $\left[B_{min}^{max}(x)\right]^{*0} = \delta(x)$.

It is sometimes more convenient to express the $n^{th}$ convolution of a box function in terms of box functions themselves rather than signum functions. The above expression consists of $n+1$ pieces, each of which can be written as the product of a polynomial in $x$ and a box function. At the $j^{th}$ boundary between these pieces ($1 \leq j \leq n$), the first $j$ signum functions will be positive and the remaining $n - j$ signum functions will be negative. Thus, we can write

$$\left[B_{min}^{max}(x)\right]^{*n} = \frac{1}{2(n-1)!}\sum_{j=1}^{n}\left[\sum_{k=0}^{j-1}(-1)^k\binom{n}{k}(x - k\,max - (n-k)\,min)^{n-1} - \sum_{k=j}^{n}(-1)^k\binom{n}{k}(x - k\,max - (n-k)\,min)^{n-1}\right]B_{n\,min+(j-1)(max-min)}^{n\,min+j(max-min)}(x) \quad (A.1)$$

*A.3 Derivation of $_R\Lambda_m^n(x)$ and $_L\Lambda_m^n(x)$*

Using Equation (A.1), we can write expressions of the form $\left[B_{w_{min}}^{w_{max}}(x)\right]^{*n} * \left[B_{w_{min}}^{0}(x)\right]^{*m}$ as

$$_R\Lambda_m^n(x) = \left[B_{w_{min}}^{w_{max}}(x)\right]^{*n} * \left[B_{w_{min}}^{0}(x)\right]^{*m}$$

$$= \left[\frac{1}{2(n-1)!}\sum_{h=1}^{n}\underbrace{\left[\sum_{k=0}^{h-1}(-1)^k\binom{n}{k}(x - k\,w_{max} - (n-k)w_{min})^{n-1} - \sum_{k=h}^{n}(-1)^k\binom{n}{k}(x - k\,w_{max} - (n-k)w_{min})^{n-1}\right]}_{P_h^{n-1}(x)}B_{n\,w_{min}+(h-1)(w_{max}-w_{min})}^{n\,w_{min}+h(w_{max}-w_{min})}(x)\right] *$$

$$\left[\frac{1}{2(m-1)!}\sum_{j=1}^{m}\underbrace{\left[\sum_{k=0}^{j-1}(-1)^k\binom{U}{k}(x - (m-k)w_{min})^{m-1} - \sum_{k=j}^{m}(-1)^k\binom{m}{k}(x - (m-k)w_{min})^{m-1}\right]}_{_RQ_j^{m-1}(x)}B_{(m-j+1)w_{min}}^{(m-j)w_{min}}(x)\right]$$

$$= \frac{1}{4(n-1)!(m-1)!}\left(\left[\sum_{h=1}^{n}P_h^{n-1}(x)B_{n\,w_{min}+(h-1)(w_{max}-w_{min})}^{n\,w_{min}+h(w_{max}-w_{min})}(x)\right] * \left[\sum_{j=1}^{m}{_RQ_j^{m-1}}(x)B_{(m-j+1)w_{min}}^{(m-j)w_{min}}(x)\right]\right)$$

$$= \frac{1}{4(n-1)!(m-1)!}\sum_{h=1}^{n}\sum_{j=1}^{m}\left[P_h^{n-1}(x)B_{n\,w_{min}+(h-1)(w_{max}-w_{min})}^{n\,w_{min}+h(w_{max}-w_{min})}(x)\right] * \left[{_RQ_j^{m-1}}(x)B_{(m-j+1)w_{min}}^{(m-j)w_{min}}(x)\right]$$

where $P_h^t(x)$ and $_RQ_j^t(x)$ are the $t^{th}$-order polynomials in $x$ obtained by grouping the coefficients of the box functions. We can use the binomial theorem and some summation gymnastics to write $P_h^t(x)$ as a sum of coefficients $p_r^t$ times powers of $x$

$$P_h^t(x) = \left[\sum_{k=0}^{h-1}(-1)^k\binom{t+1}{k}(x - k\,w_{max} - (t-k+1)w_{min})^t\right] - \left[\sum_{k=h}^{t+1}(-1)^k\binom{t+1}{k}(x - k\,w_{max} - (t-k+1)w_{min})^t\right]$$

$$= \left[\sum_{k=0}^{h-1}(-1)^k\binom{t+1}{k}\left(\sum_{r=0}^{t}\binom{t}{r}(-k\,w_{max} - (t-k+1)w_{min})^{t-r}x^r\right)\right] - \left[\sum_{k=h}^{t+1}(-1)^k\binom{t+1}{k}\left(\sum_{r=0}^{t}\binom{t}{r}(-k\,w_{max} - (t-k+1)w_{min})^{t-r}x^r\right)\right]$$

$$= \left[\sum_{k=0}^{h-1}\sum_{r=0}^{t}(-1)^k\binom{t+1}{k}\binom{t}{r}(-k\,w_{max} - (t-k+1)w_{min})^{t-r}x^r\right] - \left[\sum_{k=h}^{t+1}\sum_{r=0}^{t}(-1)^k\binom{t+1}{k}\binom{t}{r}(-k\,w_{max} - (t-k+1)w_{min})^{t-r}x^r\right]$$

$$= \left[\sum_{r=0}^{t}\sum_{k=0}^{h-1}(-1)^k\binom{t+1}{k}\binom{t}{r}(-k\,w_{max} - (t-k+1)w_{min})^{t-r}x^r\right] - \left[\sum_{r=0}^{t}\sum_{k=h}^{t+1}(-1)^k\binom{t+1}{k}\binom{t}{r}(-k\,w_{max} - (t-k+1)w_{min})^{t-r}x^r\right]$$

$$= \sum_{r=0}^{t}\binom{t}{r}\underbrace{\left[\sum_{k=0}^{h-1}(-1)^k\binom{t+1}{k}(-k\,w_{max} - (t-k+1)w_{min})^{t-r} - \sum_{k=h}^{t+1}(-1)^k\binom{t+1}{k}(-k\,w_{max} - (t-k+1)w_{min})^{t-r}\right]}_{p_r^t}x^r$$

where we define $0^0 \equiv 1$ so that the binomial expansion $(x+a)^n = \sum_{k=0}^{n}\binom{n}{k}a^{n-k}x^k$ works even when $a = 0$. Likewise, $_RQ_j^t(x)$ can be written as a sum of coefficients $_Rq_s^t$ times powers of $x$

$$_RQ_j^t(x) = \sum_{s=0}^{t}\binom{t}{s}\underbrace{\left[\sum_{k=0}^{j-1}(-1)^k\binom{t+1}{k}(-(t-k+1)w_{min})^{t-s} - \sum_{k=j}^{t+1}(-1)^k\binom{t+1}{k}(-(t-k+1)w_{min})^{t-s}\right]}_{_Rq_s^t}x^s$$

Thus, we can write $_R\Lambda_m^n(x)$ as

$$_R\Lambda_m^n(x) = \frac{1}{4(n-1)!(m-1)!}\sum_{h=1}^{n}\sum_{j=1}^{m}\left[P_h^{n-1}(x)B_{n\,w_{min}+(h-1)(w_{max}-w_{min})}^{n\,w_{min}+h(w_{max}-w_{min})}(x)\right] * \left[_RQ_j^{m-1}(x)B_{(m-j+1)w_{min}}^{(m-j)w_{min}}(x)\right]$$

$$= \frac{1}{4(n-1)!(m-1)!}\sum_{h=1}^{n}\sum_{j=1}^{m}\left[\left(\sum_{r=0}^{n-1}p_r^{n-1}x^r\right)B_{n\,w_{min}+(h-1)(w_{max}-w_{min})}^{n\,w_{min}+h(w_{max}-w_{min})}(x)\right] * \left[\left(\sum_{s=0}^{m-1}{}_Rq_s^{m-1}x^s\right)B_{(m-j+1)w_{min}}^{(m-j)w_{min}}(x)\right] \quad (A.2)$$

$$= \frac{1}{4(n-1)!(m-1)!}\sum_{h=1}^{n}\sum_{j=1}^{m}\sum_{r=0}^{n-1}\sum_{s=0}^{m-1}p_r^{n-1}\,{}_Rq_s^{m-1}\left[x^r B_{n\,w_{min}+(h-1)(w_{max}-w_{min})}^{n\,w_{min}+h(w_{max}-w_{min})}(x) * x^s B_{(m-j+1)w_{min}}^{(m-j)w_{min}}(x)\right]$$

$$= \frac{1}{4(n-1)!(m-1)!}\sum_{h=1}^{n}\sum_{j=1}^{m}\sum_{r=0}^{n-1}\sum_{s=0}^{m-1}p_r^{n-1}\,{}_Rq_s^{m-1}\,{}_s^r\Omega_{[(m-j+1)w_{min},\,(m-j)w_{min}]}^{[n\,w_{min}+(h-1)(w_{max}-w_{min}),\,n\,w_{min}+h(w_{max}-w_{min})]}(x)$$

where ${}_m^n\Omega_{[c,d]}^{[a,b]}(x)$ is defined in Section A.4. Note that we define ${}_R\Lambda_m^0(x) \equiv \left[B_{w_{\min}}^0(x)\right]^{*m}$, ${}_R\Lambda_0^n(x) \equiv \left[B_{w_{\min}}^{w_{\max}}(x)\right]^{*n}$ and ${}_R\Lambda_0^0(x) \equiv \delta(x)$.

A similar derivation for ${}_L\Lambda_m^n(x)$ gives

$$ {}_L\Lambda_m^n(x) = \frac{1}{4(n-1)!(m-1)!} \sum_{h=1}^{n}\sum_{j=1}^{m}\sum_{r=0}^{n-1}\sum_{s=0}^{m-1} p_r^{n-1}\, {}_L q_s^{m-1}\, {}_s^r\Omega_{[(j-1)w_{\max},\ jw_{\max}]}^{[nw_{\min}+(h-1)(w_{\max}-w_{\min}),\ nw_{\min}+h(w_{\max}-w_{\min})]}(x) $$

where

$$ {}_L q_s^t = \binom{t}{s}\left[\sum_{k=0}^{j-1}(-1)^k\binom{t+1}{k}(-k\,w_{\max})^{t-s} - \sum_{k=j}^{t+1}(-1)^k\binom{t+1}{k}(-k\,w_{\max})^{t-s}\right] $$

*A.4 Derivation of ${}_m^n\Omega_{[c,d]}^{[a,b]}(x)$*

In order to complete our derivations for ${}_R\Lambda_m^n(x)$ and ${}_L\Lambda_m^n(x)$ above, we must show how to evaluate expressions of the form

$$ {}_m^n\Omega_{[c,d]}^{[a,b]}(x) \equiv x^n B_a^b(x) * x^m B_c^d(x) = \int_{-\infty}^{\infty} v^n B_a^b(x)(x-v)^m B_c^d(x)\,dv $$

Since $n,m \in \mathbb{N}$, $n,m \geq 0$, $x,a,b,c,d \in \mathbb{R}$ and both $a \leq b$ and $c \leq d$, there are five cases to consider:

1) $B_c^d(x-v)$ is completely to the left of $B_a^b(v)$
2) The right edge of $B_c^d(x-v)$ overlaps the left edge of $B_a^b(v)$
3) $B_c^d(x-v)$ and $B_a^b(v)$ overlap completely
4) The right edge of $B_c^d(x-v)$ extends beyond the right edge of $B_a^b(v)$
5) $B_c^d(x-v)$ is completely to the right of $B_a^b(v)$

Taking into account these different regions and the overlap integrals that occur in each case, we have

$$_m^n\Omega_{[c,d]}^{[a,b]}(x) = \begin{cases} \begin{cases} 0 & x < a+c \\ \int_a^{x-c} v^n(x-v)^m\,dv & a+c \le x \le a+d \\ \int_{x-d}^{x-c} v^n(x-v)^m\,dv & a+d \le x \le b+c \\ \int_{x-d}^{b} v^n(x-v)^m\,dv & b+c \le x \le b+d \\ 0 & x > b+d \end{cases} & b-a \ge d-c \\[2em] \begin{cases} 0 & x < a+c \\ \int_a^{x-c} v^n(x-v)^m\,dv & a+c \le x \le a+d \\ \int_a^{b} v^n(x-v)^m\,dv & a+d \le x \le b+c \\ \int_{x-d}^{b} v^n(x-v)^m\,dv & b+c \le x \le b+d \\ 0 & x > b+d \end{cases} & d-c \ge b-a \end{cases}$$

The integral can be evaluated indefinitely and then the various limits of integration can be substituted. There are four cases to consider:

1) When $n = m = 0$: $\int dv = v$

2) When $n = 0$ but $m > 0$: $\int (x-v)^m\,dv = -\dfrac{(x-v)^{m+1}}{m+1}$

3) When $m = 0$ but $n > 0$: $\int v^n\,dv = \dfrac{v^{n+1}}{n+1}$

4) When both $n, m > 0$: $\int v^n(x-v)^m\,dv = \dfrac{v^{n+1}(x-v)^m\, {}_2F_1\!\left(n+1,-m;n+2;\dfrac{v}{x}\right)}{(n+1)\left(1-\dfrac{v}{x}\right)^m}$

Here ${}_2F_1(a,b;c;x)$ is Gauss' hypergeometric function. In addition, case (4) has a singularity at $v = x$, so we must take a limit there:

$$\lim_{v \to x} \frac{v^{n+1}(x-v)^m \, _2F_1\left(n+1,-m;n+2;\dfrac{v}{x}\right)}{(n+1)\left(1-\dfrac{v}{x}\right)^m} = \frac{(-1)^m x^{n+1} \Gamma(n+2)\Gamma(m+1)}{(n+1)\left(-\dfrac{1}{x}\right)^m m\,\Gamma(n+m+2)}$$

where $\Gamma(x)$ is Euler's gamma function.

Thus, if we define

$$h(v) \equiv \begin{cases} v & n = m = 0 \\[4pt] -\dfrac{(x-v)^{m+1}}{m+1} & n = 0, m > 0 \\[4pt] \dfrac{v^{n+1}}{n+1} & m = 0, n > 0 \\[4pt] \dfrac{v^{n+1}(x-v)^m \, _2F_1\left(n+1,-m;n+2;\dfrac{v}{x}\right)}{(n+1)\left(1-\dfrac{v}{x}\right)^m} & n,m > 0, v \neq x \\[4pt] \dfrac{(-1)^m x^{n+1}\Gamma(n+2)\Gamma(m+1)}{(n+1)\left(-\dfrac{1}{x}\right)^m m\,\Gamma(n+m+2)} & n,m > 0, v = x \end{cases}$$

then we can write ${}_m^n\Omega_{[c,d]}^{[a,b]}(x)$ as

$${}_m^n\Omega_{[c,d]}^{[a,b]}(x) = \begin{cases} 0 & x < a+c \text{ or } x > b+d \\ h(x-c) - h(a) & a+c \leq x \leq a+d \\ h(b) - h(x-d) & b+c \leq x \leq b+d \\ h(x-c) - h(x-d) & b-a \geq d-c \text{ and } a+d \leq x \leq b+c \\ h(b) - h(a) & d-c \geq b-a \text{ and } a+d \leq x \leq b+c \end{cases} \quad (A.3)$$

*A.5 Derivation of ${}_R\chi^{U,A}$ and ${}_L\chi^{U,A}$*

Since they involve finite sums of products of functions and box functions, $\rho_R^{U,A}(x)$ and $\rho_L^{U,A}(x)$ are only nonzero over a limited range. This means that we can restrict the range of the numerical integration that we have to perform to be much smaller than the $\pm\infty$ shown in Equation (3.2). In order to calculate these limits for $\rho_R^{U,A}(x)$, we must find the smallest lower bound and the largest upper bound for the box functions appearing in Equation (5.1). Recall that we have assumed throughout this paper that $w_{min} \leq 0$ and $w_{max} \geq 0$. From Equation (A.1), we can determine that the first term in Equation (5.1) is nonzero only when $x \in [U w_{min}, U w_{max}]$. From Equations (A.2) and (A.3), we can determine that the second term in Equation (5.1) is nonzero only when $x \in [(U+A)w_{min}, (A-1)w_{min} + U w_{max}]$. Thus, we can conclude that $\rho_R^{U,A}(x)$ is only nonzero when $x \in [(U+A)w_{min}, U w_{max}]$. Similar considerations apply to $\rho_L^{U,A}(x)$, giving a nonzero range of $[U w_{min}, (U+A) w_{max}]$.

Finally, there are two degenerate cases that we must deal with. First, when $U = A = 0$, there are no coupling weights and thus $\rho_R^{0,0}(x) = \delta(x)$. In this case, $_R\chi^{U,A} = \int_{-\infty}^{\infty} F_R(x) \delta(x) dx = F_R(0)$. Second, when $U = 0$ but $A > 0$, $\rho_R^{0,A}(x) = {_R\rho_2^A}(x)$. Unfortunately, $_R\rho_2^A(x)$ contains a $\delta(x)$ term which will not be handled properly by numerical integration. In this case, we must write

$$_R\chi^{U,A} = \int_{(U+A)w_{min}}^{U w_{max}} F_R(x) \rho_R^{0,A}(x) dx$$

$$= \int_{(U+A)w_{min}}^{U w_{max}} F_R(x) {_R\rho_2^A}(x) dx$$

$$= \int_{(U+A)w_{min}}^{U w_{max}} F_R(x) \left[ \sum_{j=0}^{A} \binom{A}{j} (1-b_R)^j \left(\frac{b_R}{|w_{min}|}\right)^{A-j} \left[B_{w_{min}}^0(x)\right]^{*(A-j)} \right] dx$$

$$= \int_{(U+A)w_{min}}^{U w_{max}} F_R(x) \left[ \sum_{j=0}^{A-1} \binom{A}{j} (1-b_R)^j \left(\frac{b_R}{|w_{min}|}\right)^{A-j} \left[B_{w_{min}}^0(x)\right]^{*(A-j)} + (1-b_R)^A \delta(x) \right] dx$$

$$= \int_{(U+A)w_{min}}^{U w_{max}} F_R(x)(1-b_R)^A \delta(x) dx + \int_{(U+A)w_{min}}^{U w_{max}} F_R(x) \left[ \sum_{j=0}^{A-1} \binom{A}{j} (1-b_R)^j \left(\frac{b_R}{|w_{min}|}\right)^{A-j} \left[B_{w_{min}}^0(x)\right]^{*(A-j)} \right] dx$$

$$= (1-b_R)^A F_R(0) + \int_{(U+A)w_{min}}^{U w_{max}} F_R(x) \left[ \sum_{j=0}^{A-1} \binom{A}{j} (1-b_R)^j \left(\frac{b_R}{|w_{min}|}\right)^{A-j} \left[B_{w_{min}}^0(x)\right]^{*(A-j)} \right] dx$$

Similar considerations apply to $_L\chi^{U,A}$, giving

$$_L\chi^{U,A} = (1-b_L)^A F_L(0) + \int_{Uw_{min}}^{(U+A)w_{max}} F_L(x) \left[ \sum_{j=0}^{A-1} \binom{A}{j}(1-b_L)^j \left(\frac{b_L}{|w_{max}|}\right)^{A-j} \left[B_0^{w_{max}}(x)\right]^{*(A-j)} \right] dx$$

when $U = 0$ and $A > 0$.

### A.6 Derivation of $_R\hat{\chi}^{U,A}$ and $_L\hat{\chi}^{U,A}$

The approximations $\hat{F}_R(x)$ and $\hat{F}_L(x)$ require the following functions

$$\hat{I}_R^{-1}(x) = \begin{cases} 2-x & x \geq -2 \\ \infty & x < -2 \end{cases} \qquad \hat{I}_L^{-1}(x) = \begin{cases} 2-x & x \leq -2 \\ -\infty & x > -2 \end{cases}$$

$$\hat{A}_R(x) \equiv \left[\hat{I}_R^{-1}(x+\theta_{max})\right]_{w_{min}^{self}}^{w_{max}^{self}} \qquad \hat{A}_L(x) \equiv \left[\hat{I}_L^{-1}(x+\theta_{max})\right]_{w_{min}^{self}}^{w_{max}^{self}}$$

$$\hat{B}_R(x) \equiv \left[\hat{I}_R^{-1}(x+\theta_{min})\right]_{w_{min}^{self}}^{w_{max}^{self}} \qquad \hat{B}_L(x) \equiv \left[\hat{I}_L^{-1}(x+\theta_{min})\right]_{w_{min}^{self}}^{w_{max}^{self}}$$

$$\hat{\Theta}_R(w) = \int \hat{I}_R(w) dw = \begin{cases} -\frac{w^2}{2} + 2w & w \leq 4 \\ 8 - 2w & w > 4 \end{cases} \qquad \hat{\Theta}_L(w) = \int \hat{I}_L(w) dw = \begin{cases} -2w & w \leq 4 \\ -\frac{w^2}{2} + 2w - 8 & w > 4 \end{cases}$$

The approximations $\hat{\rho}_R^{U,A}(x)$ and $\hat{\rho}_L^{U,A}(x)$ require the Gaussian function with mean $\mu$ and variance $\sigma^2$ as $g(x;\mu,\sigma) \equiv e^{-\frac{(x-\mu)^2}{2\sigma^2}} / \sqrt{2\pi}\sigma$, from which we can easily calculate the following integrals

$$_RG(x) \equiv \int g(x;\mu_R,\sigma_R) dx = \frac{1}{2}\text{Erf}\left(\frac{x-\mu_R}{\sqrt{2}\sigma_R}\right)$$

$$_RG_x(x) \equiv \int x g(x;\mu_R,\sigma_R) dx = \mu_R G(x) - \sigma_R^2 g(x;\mu_R,\sigma_R)$$

$$_RG_{x^2}(x) \equiv \int x^2 g(x;\mu_R,\sigma_R) dx = \left(\mu_R^2 + \sigma_R^2\right) G(x) - \sigma_R^2 (x+\mu_R) g(x;\mu_R,\sigma_R)$$

where $\mu_R \equiv \mu_1 + {_R\mu_2}$ and $\sigma_R^2 \equiv \sigma_1^2 + {_R\sigma_2^2}$.

We can then use these expressions to define

$$G_{\hat{A}_R} \equiv \int_{-\infty}^{\infty} \hat{A}_R(x) g(x; \mu_R, \sigma_R) dx$$
$$= \left(w_{max}^{self} + \theta_{max} - 2\right){}_R G\left({}_R\beta_{max}\right) - \left(w_{min}^{self} + \theta_{max} - 2\right){}_R G\left({}_R\gamma_{max}\right) + \left({}_R G_x\left({}_R\beta_{max}\right) - {}_R G_x\left({}_R\gamma_{max}\right)\right) + \frac{1}{2}\left(w_{max}^{self} + w_{min}^{self}\right)$$

$$G_{\hat{B}_R} \equiv \int_{-\infty}^{\infty} \hat{B}_R(x) g(x; \mu_R, \sigma_R) dx$$
$$= \left(w_{max}^{self} + \theta_{min} - 2\right){}_R G\left({}_R\beta_{min}\right) - \left(w_{min}^{self} + \theta_{min} - 2\right){}_R G\left({}_R\gamma_{min}\right) + \left({}_R G_x\left({}_R\beta_{min}\right) - {}_R G_x\left({}_R\gamma_{min}\right)\right) + \frac{1}{2}\left(w_{max}^{self} + w_{min}^{self}\right)$$

$$G_{x\hat{A}_R} \equiv \int_{-\infty}^{\infty} x \hat{A}_R(x) g(x; \mu_R, \sigma_R) dx$$
$$= w_{max}^{self}\left({}_R G_x\left({}_R\beta_{max}\right) + \frac{\mu_R}{2}\right) + \left(2 - \theta_{max}\right)\left({}_R G_x\left({}_R\gamma_{max}\right) - {}_R G_x\left({}_R\beta_{max}\right)\right) - \left({}_R G_{x^2}\left({}_R\gamma_{max}\right) - {}_R G_{x^2}\left({}_R\beta_{max}\right)\right)$$
$$+ w_{min}^{self}\left(\frac{\mu_R}{2} - {}_R G_x\left({}_R\gamma_{max}\right)\right)$$

$$G_{x\hat{B}_R} \equiv \int_{-\infty}^{\infty} x \hat{B}_R(x) g(x; \mu_R, \sigma_R) dx$$
$$= w_{max}^{self}\left({}_R G_x\left({}_R\beta_{min}\right) + \frac{\mu_R}{2}\right) + \left(2 - \theta_{min}\right)\left({}_R G_x\left({}_R\gamma_{min}\right) - {}_R G_x\left({}_R\beta_{min}\right)\right) - \left({}_R G_{x^2}\left({}_R\gamma_{min}\right) - {}_R G_{x^2}\left({}_R\beta_{min}\right)\right)$$
$$+ w_{min}^{self}\left(\frac{\mu_R}{2} - {}_R G_x\left({}_R\gamma_{min}\right)\right)$$

$$G_{\hat{\Theta}_R(\hat{A}_R)} \equiv \int_{-\infty}^{\infty} \hat{\Theta}_R\left(\hat{A}_R(x)\right) g(x; \mu_R, \sigma_R) dx$$
$$= \begin{cases} 2w_{max}^{self} - \dfrac{\left(w_{max}^{self}\right)^2}{2} & w_{max}^{self} \leq 4 \\ 8 - 2w_{max}^{self} & w_{max}^{self} > 4 \end{cases} \left({}_R G\left({}_R\beta_{max}\right) + \frac{1}{2}\right) - \frac{1}{2}\left({}_R G_{x^2}\left({}_R\gamma_{max}\right) - {}_R G_{x^2}\left({}_R\beta_{max}\right)\right)$$
$$- \theta_{max}\left({}_R G_x\left({}_R\gamma_{max}\right) - {}_R G_x\left({}_R\beta_{max}\right)\right) + \left(2 - \frac{\theta_{max}^2}{2}\right)\left({}_R G\left({}_R\gamma_{max}\right) - {}_R G\left({}_R\beta_{max}\right)\right)$$
$$+ \left(2w_{min}^{self} - \frac{\left(w_{min}^{self}\right)^2}{2}\right)\left(\frac{1}{2} - {}_R G\left({}_R\gamma_{max}\right)\right)$$

$$G_{\hat{\Theta}_R(\hat{B}_R)} \equiv \int_{-\infty}^{\infty} \hat{\Theta}_R\left(\hat{B}_R(x)\right) g(x;\mu_R,\sigma_R) dx$$

$$= \begin{cases} 2w_{max}^{self} - \dfrac{\left(w_{max}^{self}\right)^2}{2} & w_{max}^{self} \leq 4 \\ 8 - 2w_{max}^{self} & w_{max}^{self} > 4 \end{cases} \left({}_R G({}_R\beta_{min}) + \dfrac{1}{2}\right) - \dfrac{1}{2}\left({}_R G_{x^2}({}_R\gamma_{min}) - {}_R G_{x^2}({}_R\beta_{min})\right)$$

$$- \theta_{min}\left({}_R G_x({}_R\gamma_{min}) - {}_R G_x({}_R\beta_{min})\right) + \left(2 - \dfrac{\theta_{min}^2}{2}\right)\left({}_R G({}_R\gamma_{min}) - {}_R G({}_R\beta_{min})\right)$$

$$+ \left(2w_{min}^{self} - \dfrac{\left(w_{min}^{self}\right)^2}{2}\right)\left(\dfrac{1}{2} - {}_R G({}_R\gamma_{min})\right)$$

where ${}_R\beta_{max} \equiv \max(-2, 2 - w_{max}^{self}) - \theta_{max}$, ${}_R\beta_{min} \equiv \max(-2, 2 - w_{max}^{self}) - \theta_{min}$, ${}_R\gamma_{max} \equiv 2 - w_{min}^{self} - \theta_{max}$ and ${}_R\gamma_{min} \equiv 2 - w_{min}^{self} - \theta_{min}$.

With these definitions, we can derive a closed-form expression for ${}_R\hat{\chi}^{U,A}$ as follows

$${}_R\hat{\chi}^{U,A} = \int_{-\infty}^{\infty} \hat{F}_R(x)\hat{\rho}_R(x) dx$$

$$= \int_{-\infty}^{\infty} \dfrac{1}{A_T}\left[(\theta_{max} - \theta_{min})\left(\hat{A}_R(x) - w_{min}^{self}\right) + \hat{\Theta}_R\left(\hat{B}_R(x)\right) - \hat{\Theta}_R\left(\hat{A}_R(x)\right) - \left(\hat{B}_R(x) - \hat{A}_R(x)\right)(x + \theta_{min})\right] g(x;\mu_R,\sigma_R) dx$$

$$= \dfrac{1}{A_T}\left[(\theta_{max} - \theta_{min})\left(\int_{-\infty}^{\infty} \hat{A}_R(x) g(x;\mu_R,\sigma_R) dx - w_{min}^{self} \int_{-\infty}^{\infty} g(x) dx\right) + \int_{-\infty}^{\infty} \hat{\Theta}_R\left(\hat{B}_R(x)\right) g(x;\mu_R,\sigma_R) dx \right.$$

$$- \int_{-\infty}^{\infty} \hat{\Theta}_R\left(\hat{A}_R(x)\right) g(x;\mu_R,\sigma_R) dx - \int_{-\infty}^{\infty} \left(\hat{B}_R(x) - \hat{A}_R(x)\right) g(x;\mu_R,\sigma_R) dx$$

$$\left. - \int_{-\infty}^{\infty} \left(\hat{B}_R(x) - \hat{A}_R(x)\right) x\, g(x;\mu_R,\sigma_R) dx\right]$$

$$= \dfrac{1}{A_T}\left[(\theta_{max} - \theta_{min})\left(G_{\hat{A}_R} - w_{min}^{self}\right) + G_{\hat{\Theta}_R(\hat{B}_R)} - G_{\hat{\Theta}_R(\hat{A}_R)} - G_{x\hat{B}_R} + G_{x\hat{A}_R} - \theta_{min}\left(G_{\hat{B}_R} - G_{\hat{A}_R}\right)\right]$$

An analogous derivation can be given for ${}_L\hat{\chi}^{U,A}$. In this case, the functions $\hat{A}_L(x)$, $\hat{B}_L(x)$, and $\hat{\Theta}_L(\cdot)$ are defined analogously to $\hat{A}_R(x)$, $\hat{B}_R(x)$ and $\hat{\Theta}_R(\cdot)$. In addition, ${}_L G(x)$, ${}_L G_x(x)$ and ${}_L G_{x^2}(x)$ are defined analogously to ${}_R G(x)$, ${}_R G_x(x)$ and ${}_R G_{x^2}(x)$ except using $\mu_L \equiv \mu_1 + {}_L\mu_2$ and $\sigma_L^2 \equiv \sigma_1^2 + {}_L\sigma_2^2$, whereas $G_{\hat{A}_L}$, $G_{\hat{B}_L}$, $G_{x\hat{A}_L}$ and $G_{x\hat{B}_L}$ are defined analogously to $G_{\hat{A}_R}$, $G_{\hat{B}_R}$, $G_{x\hat{A}_R}$ and $G_{x\hat{B}_R}$ except using ${}_L\beta_{max} \equiv 2 - w_{max}^{self} - \theta_{max}$, ${}_L\beta_{min} \equiv 2 - w_{max}^{self} - \theta_{min}$, ${}_L\gamma_{max} \equiv -2 - \theta_{max}$ and ${}_L\gamma_{min} \equiv -2 - \theta_{min}$. Finally, $G_{\hat{\Theta}_L(\hat{A}_L)}$ and $G_{\hat{\Theta}_L(\hat{B}_L)}$ are defined as

$$G_{\hat{\Theta}_L(\hat{A}_L)} \equiv \int_{-\infty}^{\infty} \hat{\Theta}_L(\hat{A}_L(x)) g(x; \mu_L, \sigma_L) dx$$

$$= \begin{cases} -2w_{\max}^{\text{self}} & w_{\max}^{\text{self}} \leq 4 \\ -8 + 2w_{\max}^{\text{self}} - \dfrac{(w_{\max}^{\text{self}})^2}{2} & w_{\max}^{\text{self}} > 4 \end{cases} \left( {}_L G({}_L \beta_{\max}) + \dfrac{1}{2} \right) - \dfrac{1}{2} \left( {}_L G_{x^2}({}_L \gamma_{\max}) - {}_L G_{x^2}({}_R \beta_{\max}) \right)$$

$$- \theta_{\max} \left( {}_L G_x({}_L \gamma_{\max}) - {}_L G_x({}_L \beta_{\max}) \right) + \left( -6 - \dfrac{\theta_{\max}^2}{2} \right) \left( {}_L G({}_L \gamma_{\max}) - {}_L G({}_L \beta_{\max}) \right)$$

$$- 2w_{\min}^{\text{self}} \left( \dfrac{1}{2} - {}_L G({}_L \gamma_{\max}) \right)$$

$$G_{\hat{\Theta}_L(\hat{B}_L)} \equiv \int_{-\infty}^{\infty} \hat{\Theta}_L(\hat{B}_L(x)) g(x; \mu_L, \sigma_L) dx$$

$$= \begin{cases} -2w_{\max}^{\text{self}} & w_{\max}^{\text{self}} \leq 4 \\ -8 + 2w_{\max}^{\text{self}} - \dfrac{(w_{\max}^{\text{self}})^2}{2} & w_{\max}^{\text{self}} > 4 \end{cases} \left( {}_L G({}_L \beta_{\min}) + \dfrac{1}{2} \right) - \dfrac{1}{2} \left( {}_L G_{x^2}({}_L \gamma_{\min}) - {}_L G_{x^2}({}_L \beta_{\min}) \right)$$

$$- \theta_{\min} \left( {}_L G_x({}_L \gamma_{\min}) - {}_L G_x({}_L \beta_{\min}) \right) + \left( -6 - \dfrac{\theta_{\min}^2}{2} \right) \left( {}_L G({}_L \gamma_{\min}) - {}_L G({}_L \beta_{\min}) \right)$$

$$- 2w_{\min}^{\text{self}} \left( \dfrac{1}{2} - {}_L G({}_L \gamma_{\min}) \right)$$

*A.7 Derivation of $\langle u \rangle(N)$ and $\langle l \rangle(N)$*

We can calculate $\langle u \rangle(N)$, the mean location of the upper bound of the net inputs received by an element, as follows:

$$\langle u \rangle(N) = \left\langle \tilde{I}_R(w) - \min \mathcal{I}^A - \sum_{j=1}^{U} w_j \right\rangle(N)$$

$$= \langle \tilde{I}_R(w) \rangle(N) - \langle \min \mathcal{I}^A \rangle(N) - \left\langle \sum_{j=1}^{U} w_j \right\rangle(N)$$

$$= \int_{-\infty}^{\infty} \tilde{I}_R(w) \frac{B_{w_{\min}^{self}}^{w_{\max}^{self}}(w)}{w_{\max}^{self} - w_{\min}^{self}} dw - \int_{-\infty}^{\infty} x \sum_{A=0}^{N-1} {}_R\rho_2^A(x) P(A|N-1) dx - \int_{-\infty}^{\infty} x \sum_{U=0}^{N-1} \rho_1^U(x) P(U|N-1) dx$$

$$= \frac{1}{w_{\max}^{self} - w_{\min}^{self}} \int_{w_{\min}^{self}}^{w_{\max}^{self}} \tilde{I}_R(w) dw - \sum_{A=0}^{N-1} P(A|N-1) \int_{-\infty}^{\infty} x \, {}_R\rho_2^A(x) dx - \sum_{U=0}^{N-1} P(U|N-1) \int_{-\infty}^{\infty} x \, \rho_1^U(x) dx$$

$$= \frac{1}{w_{\max}^{self} - w_{\min}^{self}} \int_{w_{\min}^{self}}^{w_{\max}^{self}} \tilde{I}_R(w) dw - \sum_{A=0}^{N-1} {}_R\mu_2(A) P(A|N-1) - \sum_{U=0}^{N-1} \mu_1(U) P(U|N-1)$$

where we have taken advantage of the fact that we have already calculated the mean values of the distribution $\rho_1^U(x)$ of $\sum_{j=1}^{U} w_j$ and the distribution ${}_R\rho_2^A(x)$ of $\min \mathcal{I}^A$ in Section 5. Since we are only studying coupling weight ranges that are symmetric about 0, the last term drops out because $\mu_1 = 0$ for symmetric weight ranges. The first term can be expressed using $\Theta_R(w)$ and $\Theta_L(w)$ derived in Appendix A.1, and we can express $P(A|N-1)$ as $P(\tilde{\mathcal{R}}_A^{N-1})$. These considerations give the expression for $\langle u \rangle(N)$ shown in Section 9. Similar considerations apply to $\langle l \rangle(N)$.

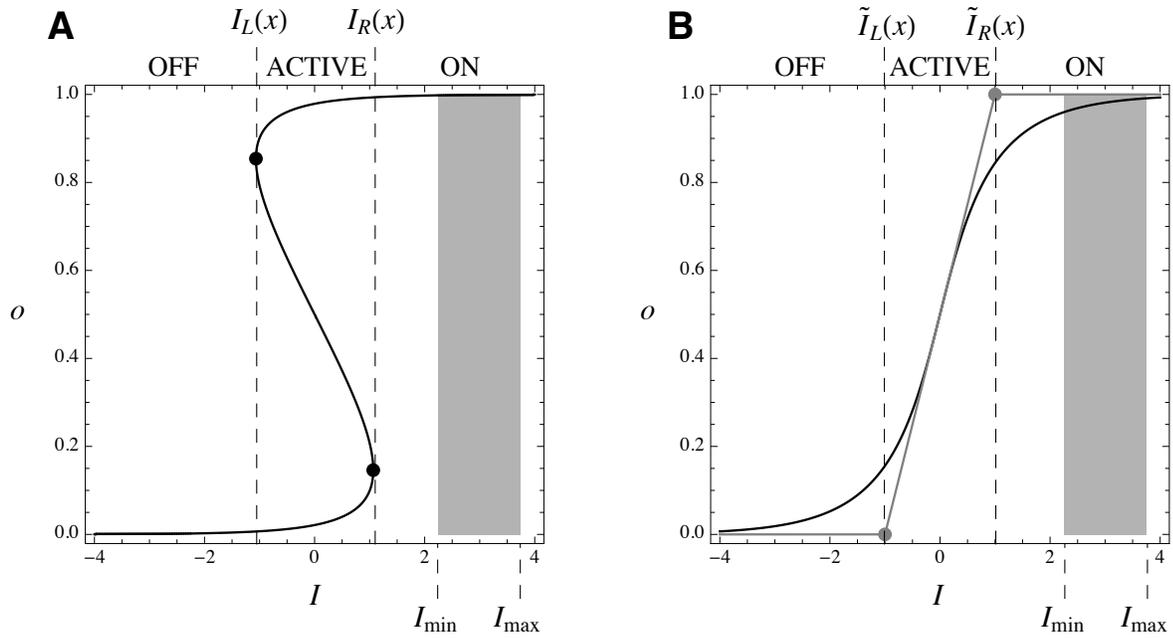

**Figure 1**: Representative steady-state input/output (SSIO) diagrams of a single CTSN for (A) $w = 8$ and (B) $w = 2$. The solid curves show the output space location of the element's equilibrium points as a function of the net input $I = J + \theta$. Note that the SSIO becomes folded for $w > 4$, indicating the existence of three equilibrium points. When the SSIO is folded, the left and right edges of the fold are given by $I_L(w)$ and $I_R(w)$, respectively (black points in A). An element whose net input is to the left of $I_L(w)$, between $I_L(w)$ and $I_R(w)$, or to the right of $I_R(w)$ will be saturated OFF, ACTIVE, or saturated ON, respectively. The range of net input is indicated by gray rectangles, whose lower and upper limits are denoted by $I_{min} = J_{min} + \theta$ and $I_{max} = J_{max} + \theta$, respectively. Thus, changing an element's bias will shift the gray rectangles accordingly. The gray line in B shows a piecewise linear SSIO approximation, which suggests using the intersections of the linear pieces (gray points) as the analogues of the fold edges in part A when $w < 4$. In this example, the relationship between the SSIO curves and the range of inputs is such that the element shown is saturated ON.

Figure 2

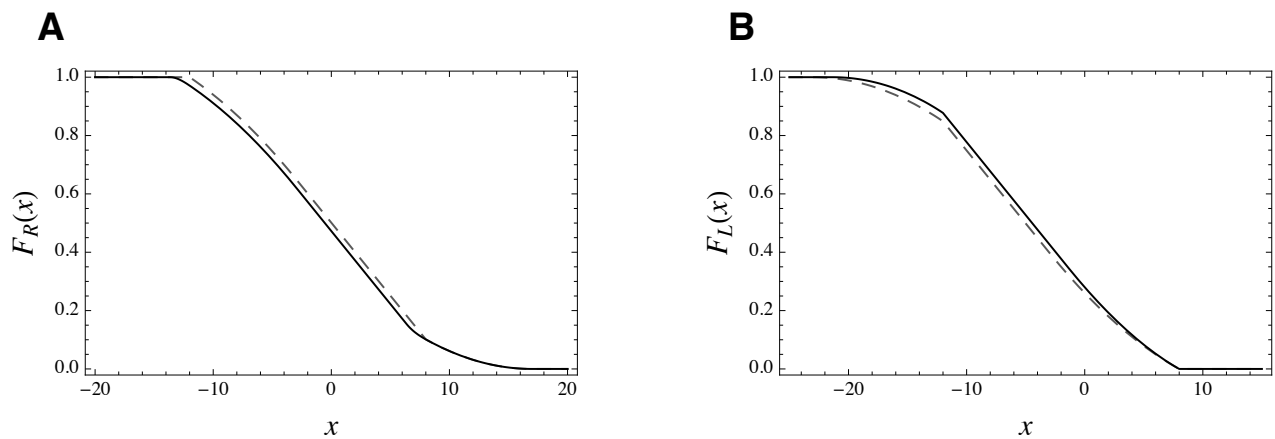

**Figure 2**: Sample plots of $F_R(x)$ (A) and $F_L(x)$ (B) for $[\theta_{min}, \theta_{min}] = [-10, 10]$ and $\left[w^{self}_{min}, w^{self}_{max}\right] = [-5, 15]$. The exact curves are shown in black and the approximations $\hat{F}_R(x)$ and $\hat{F}_L(x)$ are shown as dashed gray curves.

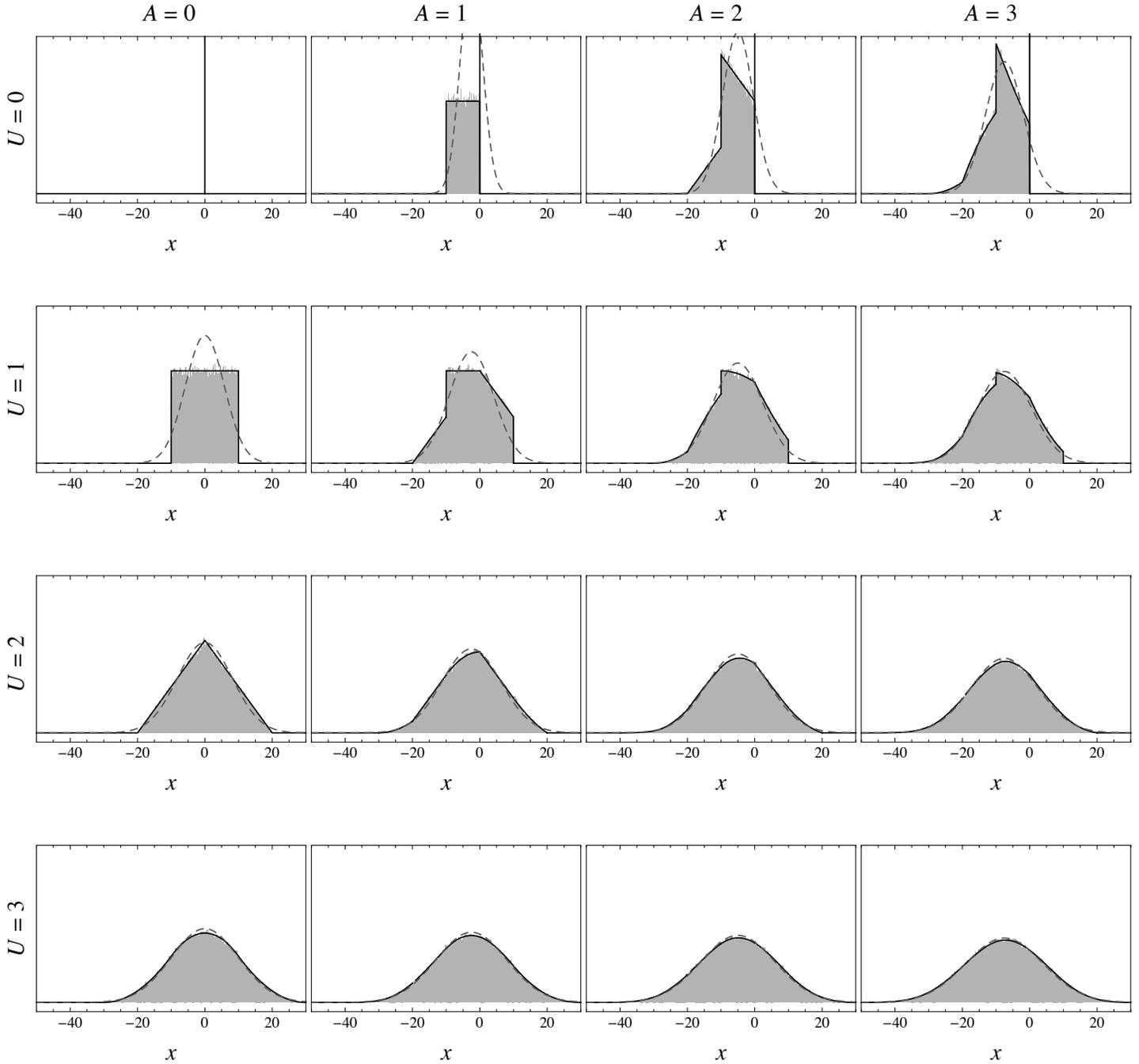

**Figure 3**: Sample plots of $\rho_R^{U,A}(x)$ show the approach to a normal distribution with increasing $U$ and $A$. Here $[w_{\min}, w_{\max}] = [-10, 10]$. Histograms drawn from $10^5$ random samples are shown in gray. The exact curve derived in Equation (5.1) is shown in black. The dashed gray curve shows the normal approximation $\hat{\rho}_R^{U,A}(x)$. All of these plots have the same vertical scale. Note the delta functions at 0 from $\rho_2^A(x)$ in the first row when $U = 0$. The approach to a normal distribution is slowest for small $U$, when most of the contributing elements are ACTIVE.

**Figure 4**

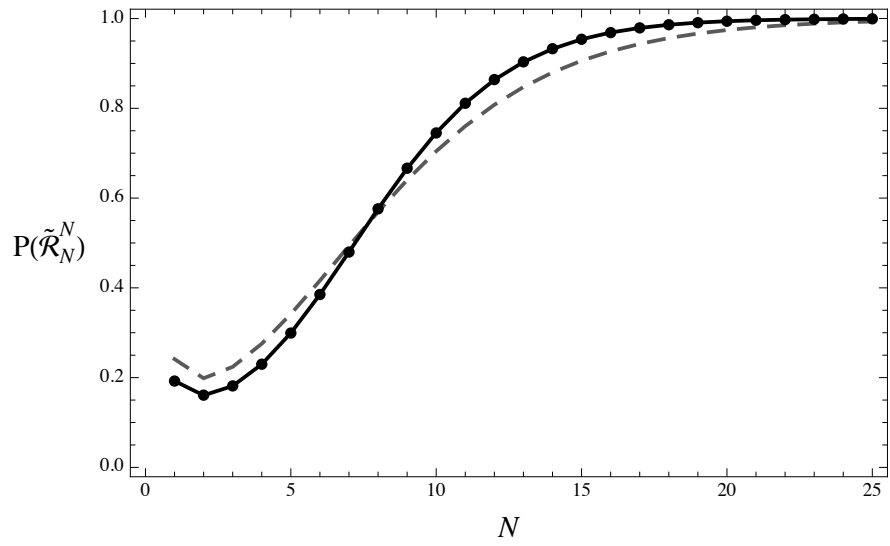

**Figure 4**: The scaling of $P(\tilde{\mathcal{R}}_N^N)$ with $N$ for $[\theta_{\min}, \theta_{\max}] = [w_{\min}, w_{\max}] = [-10, 10]$ and $\left[w_{\min}^{\text{self}}, w_{\max}^{\text{self}}\right] = [-5, 15]$. The exact curve derived in Equation (7.2) is shown in black. The black dots denote data obtained from the mean of $10^6$ random samples for each point. The gray dashed curve shows the approximation $\hat{P}(\tilde{\mathcal{R}}_N^N)$.

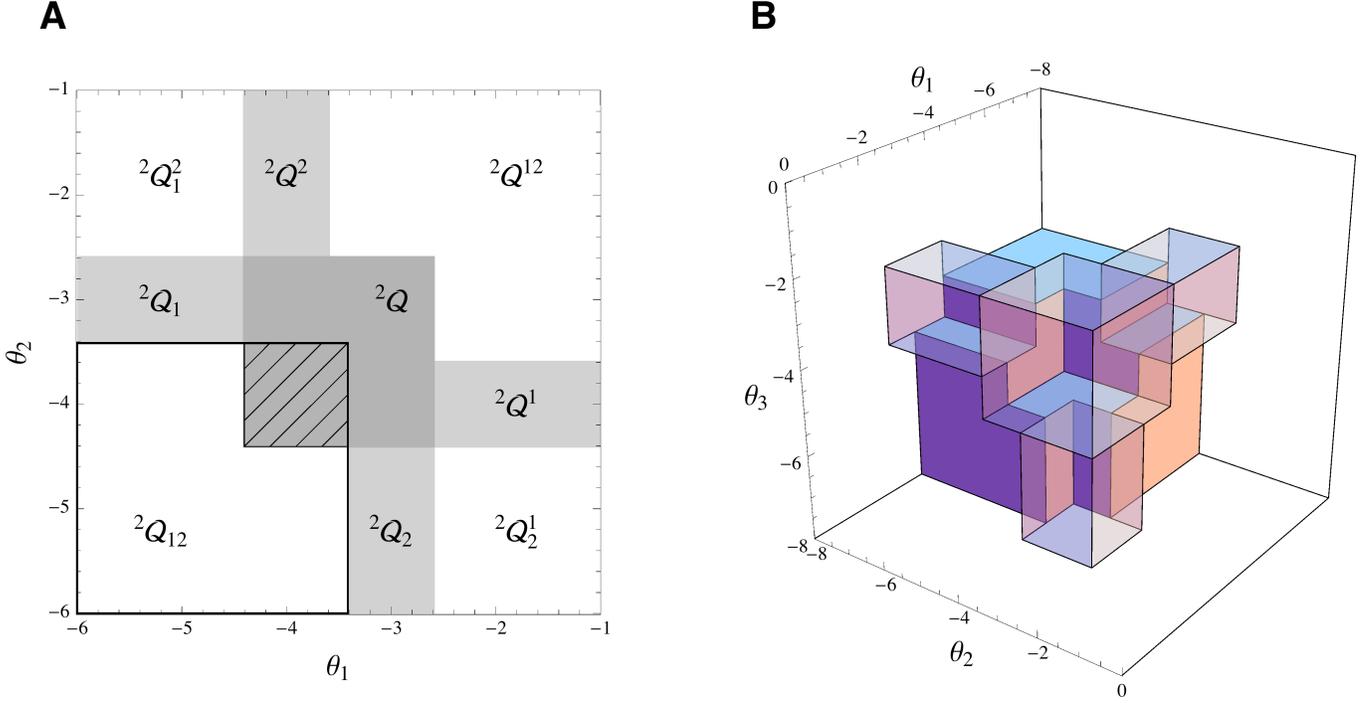

**Figure 5**: Low-dimensional illustrations of the overlap considerations that go into the derivation of Equation (7.4) for $^N S_D^U$. (A) Saturation regions in bias space for a 2-element CTSN with the fixed weight matrix $\mathbf{W} = \begin{pmatrix} 6 & 1 \\ 1 & 6 \end{pmatrix}$. $\mathcal{R}_0^2$ is shown in white, $\mathcal{R}_1^2$ is shown in light gray, and $\mathcal{R}_2^2$ is shown in dark gray. Each region consists of a disjoint union of subregions denoted by the symbol $^2\mathcal{Q}$. We focus on the nonconvex subregion $^2\mathcal{Q}_{12}$, in which both elements are saturated OFF. The large square outlined in black that surrounds $^2\mathcal{Q}_{12}$ denotes its bounding box. The cross-hatched smaller square shows the intersection between the bounding box of $^2\mathcal{Q}_{12}$ and $^2\mathcal{Q}$. This is the subregion whose probability must be subtracted from the probability of the bounding box of $^2\mathcal{Q}_{12}$ in order to obtain the probability of $^2\mathcal{Q}_{12}$. (B) The subregion $^3\mathcal{Q}_{123}$ of a 3-element CTSN with the fixed weight matrix $\mathbf{W} = \begin{pmatrix} 6 & 1 & 1 \\ 1 & 6 & 1 \\ 1 & 1 & 6 \end{pmatrix}$. $^3\mathcal{Q}_{123}$ itself is the nonconvex solid object in the background of this plot, which is missing its frontmost corner and the three adjacent edges. The transparent cube in the foreground is $^3\mathcal{Q}$, while the three transparent rectangular solids are $^3\mathcal{Q}_1$, $^3\mathcal{Q}_2$ and $^3\mathcal{Q}_3$. The probability of the intersection of the bounding cube of $^3\mathcal{Q}_{123}$ with these transparent subregions must be subtracted from the probability of the bounding cube of $^3\mathcal{Q}_{123}$ to obtain the probability of $^3\mathcal{Q}_{123}$.

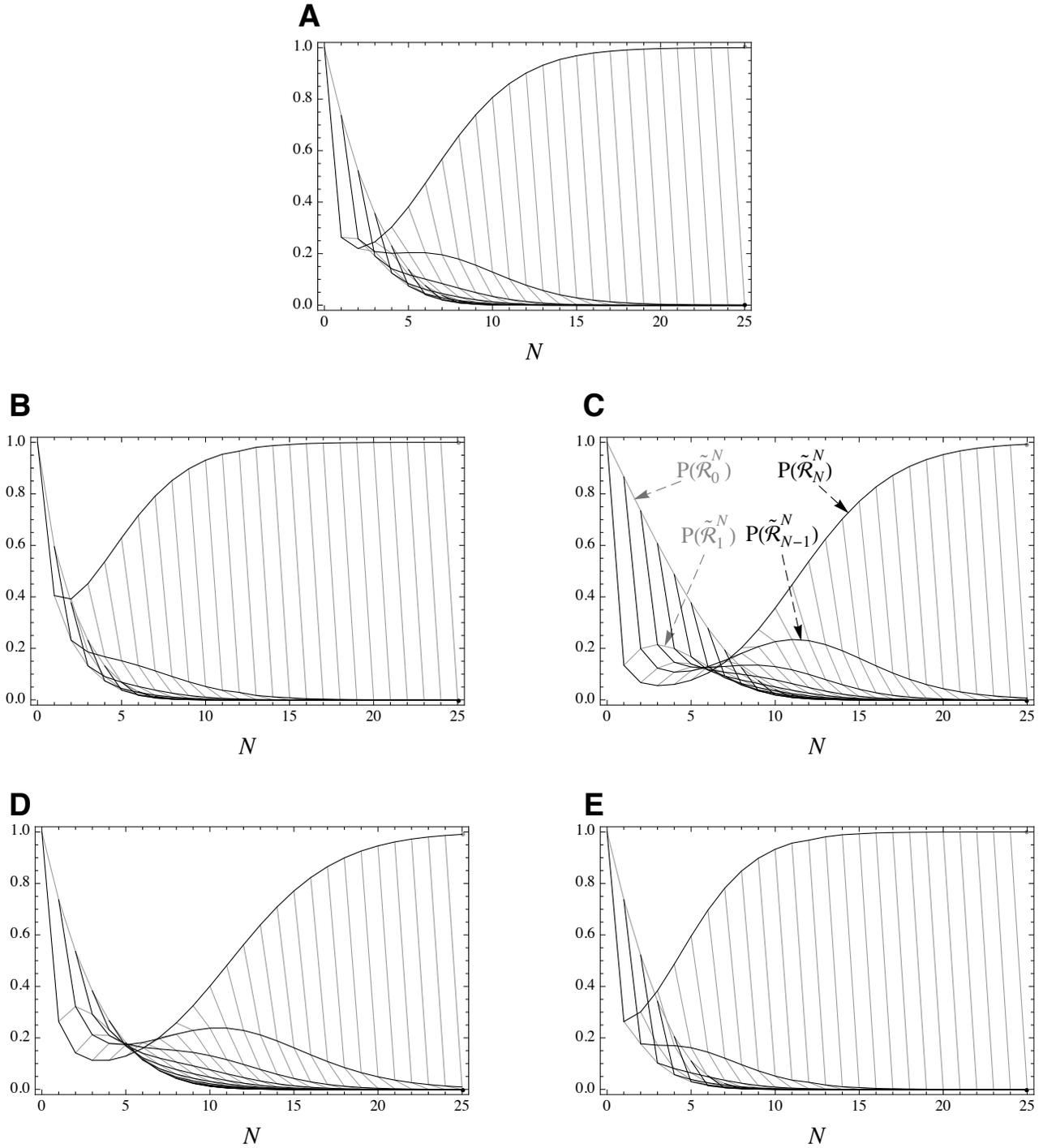

**Figure 6**: Plots of the scaling of $P(\tilde{\mathcal{R}}_M^N)$, the probability of encountering $M$-dimesnional dynamics in an $N$-element CTSN, with $N$ and $M$ for different parameter sampling ranges. $P(\tilde{\mathcal{R}}_{N-k}^N)$ is shown in black for $k$ varying from 0 to $N$. $P(\tilde{\mathcal{R}}_M^N)$ is shown in gray, with $M$ varying from 0 to $N$. (A) Nominal parameter sampling ranges $[\theta_{min}, \theta_{min}] = [w_{min}, w_{max}] = [w_{min}^{self}, w_{max}^{self}] = [-10, 10]$. (B) $[\theta_{min}, \theta_{min}] = [-5, 5]$. (C) $[\theta_{min}, \theta_{min}] = [-20, 20]$. (D) $[w_{min}, w_{max}] = [-5, 5]$. (E) $[w_{min}, w_{max}] = [-20, 20]$.

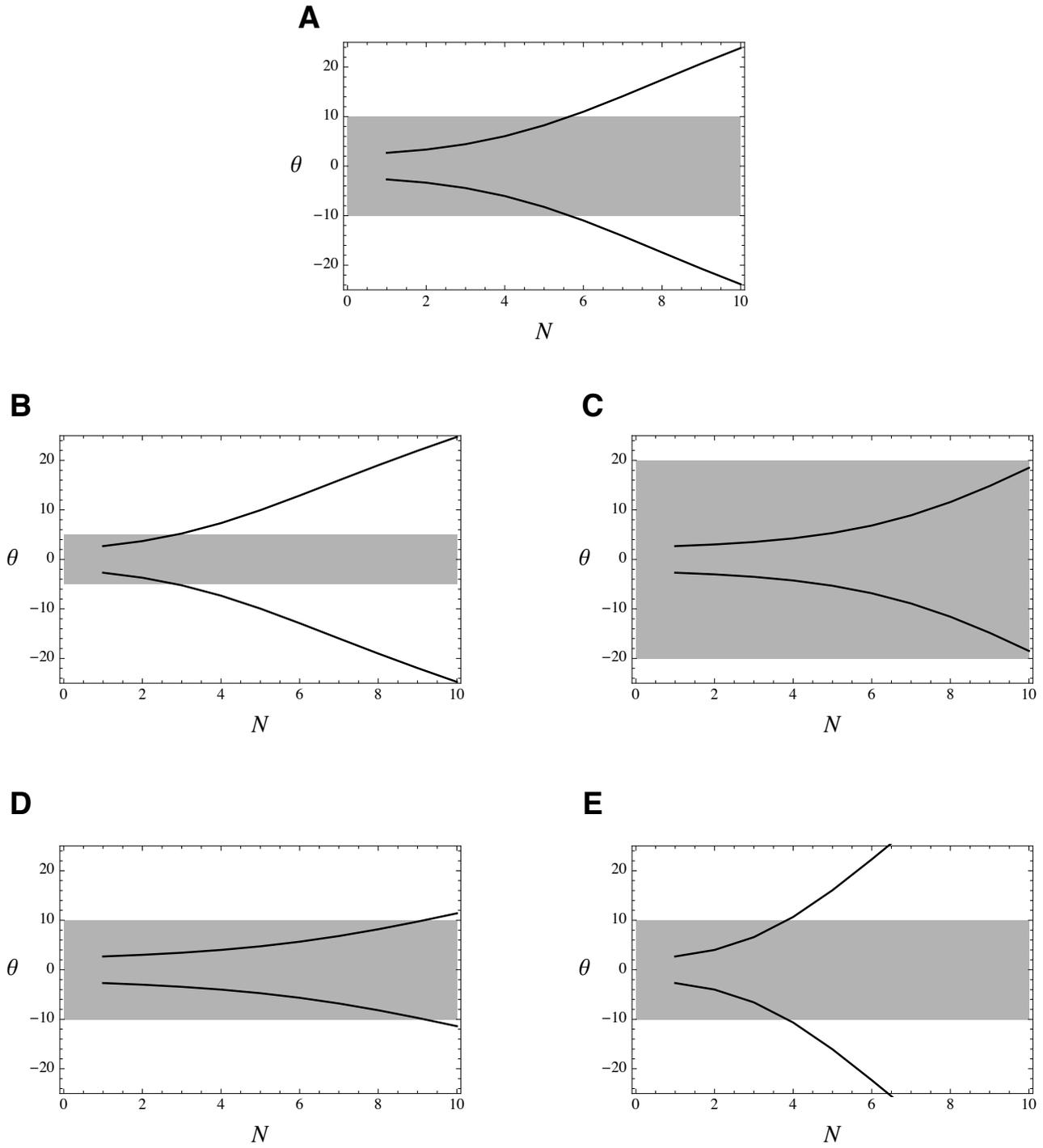

**Figure 7**: Plots of the scaling of $<u>(N)$ and $<l>(N)$ (upper and lower black curves, respectively) for an arbitrary element relative to the allowable bias range for that element (gray rectangle) for different parameter sampling ranges corresponding to those used in Figure 6. (A) Nominal parameter sampling ranges $[\theta_{min}, \theta_{min}] = [w_{min}, w_{max}] = [w_{min}^{self}, w_{max}^{self}] = [-10, 10]$. (B) $[\theta_{min}, \theta_{min}] = [-5, 5]$. (C) $[\theta_{min}, \theta_{min}] = [-20, 20]$. (D) $[w_{min}, w_{max}] = [-5, 5]$. (E) $[w_{min}, w_{max}] = [-20, 20]$.

**Figure 8**

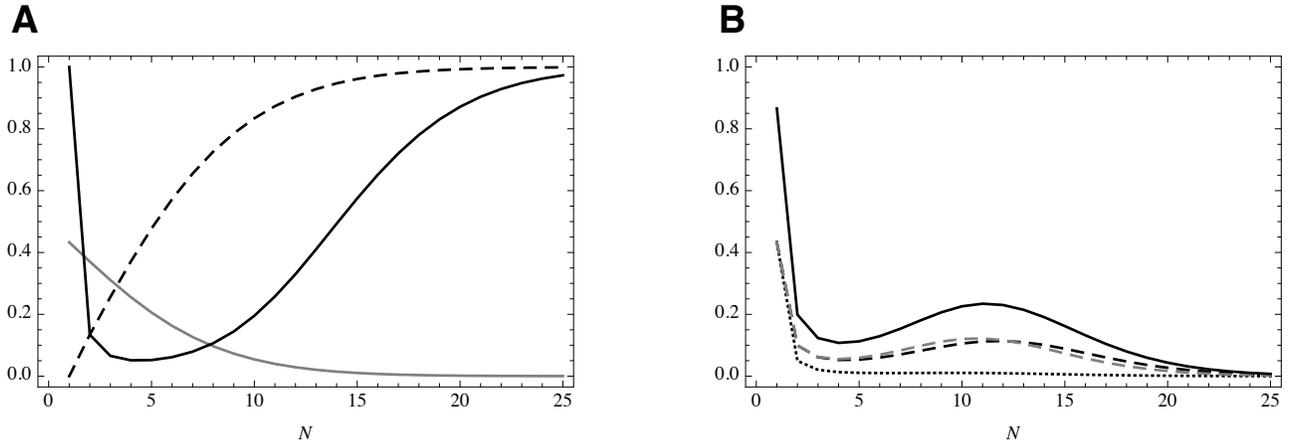

**Figure 8**: A reconstruction of the form of $P(\tilde{\mathcal{R}}_{N-1}^{N})$ for $[\theta_{min}, \theta_{min}] = [-20, 20]$ and $[w_{min}, w_{max}] = [w_{min}^{self}, w_{max}^{self}] = [-10, 10]$. (A) The mean normalized length of a region in which one element is saturated on (solid gray curve), the mean normalized width of this region (dashed black curve) and this mean normalized width raised to the (N-1)th power (solid black curve) as a function of N. (B) The product of the mean normalized length and the mean normalized width raised to the (N-1)th power (dotted curve), the product of the former curve and N (dashed black curve), the corresponding product for a region in which one element is saturated off (dashed gray curve) and the sum of the the two dashed curves (solid black curve), which corresponds to the $P(\tilde{\mathcal{R}}_{N-1}^{N})$ curve in Figure 6C.

**Figure A1**

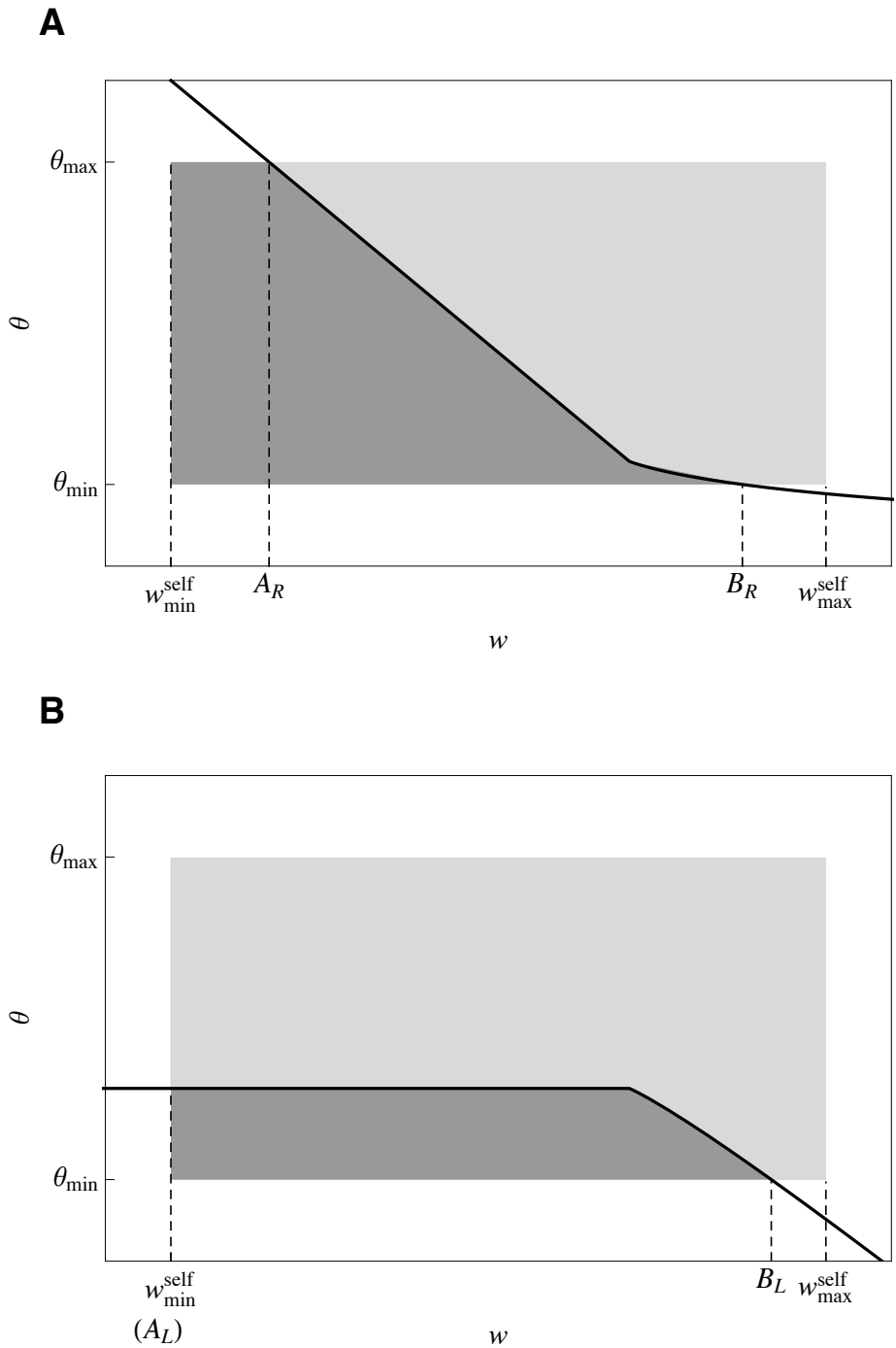

**Figure A1**: Plots of the curves $\tilde{I}_R(w) - x$ (A) and $\tilde{I}_L(w) - x$ (B) relative to the constraint rectangle $[w_{\min}^{\text{self}}, w_{\max}^{\text{self}}] \times [\theta_{\min}, \theta_{\max}]$ shown in gray. The actual area to be calculated is indicated in dark gray. In each case, $A$ denotes the intersection of the black curve with $\theta_{\max}$ (clipped to $[w_{\min}^{\text{self}}, w_{\max}^{\text{self}}]$) and $B$ denotes the intersection of the black curve with $\theta_{\min}$ (clipped to $[w_{\min}^{\text{self}}, w_{\max}^{\text{self}}]$).